\documentclass[amsmath,amssymb,aps,reprint,prb,floatfix]{revtex4-2}

\usepackage{graphicx}
\usepackage{dcolumn}
\usepackage{bm}
\usepackage[hidelinks]{hyperref}
\usepackage{upgreek} 

\begin{document}

\preprint{APS/123-QED}

\title{Computational Studies on O2-P2 Phase-Transition Dynamics in Layered-Oxide Sodium-Ion Cathode Materials}

\author{Konstantin Köster}
\author{Payam Kaghazchi}
\email{p.kaghazchi@fz-juelich.de}
\affiliation{%
Materials Synthesis and Processing (IMD-2), Institute of Energy Materials and Devices, Forschungszentrum Jülich GmbH, Wilhelm-Johnen-Straße, Jülich, 52425, Germany \\
MESA+ Institute, University of Twente, Hallenweg 15, Enschede, 7522, NH, The Netherlands
}%

\date{\today}

\begin{abstract}
Sodium-ion batteries have gained much interest over the past years and especially layered oxides are highly considered as cathodes for the next generation of batteries. However, there are still significant challenges to overcome in these materials for practical applications mainly related to capacity degradation and voltage fading. A key influence factor for these challenges are phase transitions that occur by gliding of layers during operation of these materials. Until now there is limited atomistic-level understanding on such transitions as simulations of these processes are computationally demanding. In this work, we trained a classical pairwise Coulomb-Buckingham potential versus extensive \textit{ab initio} data using a genetic algorithm to study O2-P2 phase transitions in Na\textsubscript{\textit{x}}CoO\textsubscript{2}. Our density functional theory~(DFT) and classical potential calculations show that phase transition barriers decrease upon desodiation and are further lowered if dynamic conditions are considered through molecular dynamics simulations. Our developed classical potential is able to capture phase transitions and its related increase in the Na-ion diffusivity under standard lab conditions at the $\upmu$s timescale of molecular dynamics simulation. Furthermore, it is found that the phase transition occurs gradually \textit{via} various OP\textit{n} phases.
\end{abstract}

\maketitle


\section{Introduction}
Layered oxide cathode active materials are among the most promising for sodium-ion batteries (SIBs) as they are already successfully applied in the widely commercialized lithium-ion batteries (LIBs). While LIBs are superior over SIBs in terms of total capacity and capacity retention, SIBs can offer several advantages over LIBs when ecological and economical/strategic perspectives are considered. The high, widely-distributed abundance of Na compared to Li can yield strategic as well as cost benefits while in addition often less critical metals are required in SIBs compared to LIBs. Despite these considerable advantages of SIBs over LIBs their wide application is still hindered by there inferior electrochemical performance. Key challenges in improving the performance of SIBs are often connected to the much larger ionic radius of sodium compared to lithium which causes more drastic changes in the host materials upon (de-)intercalation during (dis-)charging. More precisely, for the aforementioned layered oxide cathode materials often phase transitions are observed during operation~\cite{Wang.2024, Yang.2024, Wang.2018, Hwang.2017, Zhang.2024}. The phases can be classified by the sodium-ion coordination-environment such as octahedral~(O) and and prismatic~(P) and the stacking periodicity of the layers along the \textit{c}-direction~\cite{Delmas.1980} while the most commonly observed phases are P2 (intermediate sodium concentration) and O3 (high sodium concentration). However, during operation of the cathode material the varying sodium concentration often yields to (ir-)reversible phase transitions. The most important transitions are P2 to O2 and O3 to P3 transitions while also transitions to mixed stackings such as OP2 or OP4 are observed~\cite{Hwang.2017, Zhang.2024}. 

Whilst there is a common understanding that these transitions are caused by gliding of transition-metal oxygen slabs, there are little mechanistic insights until now as this requires atomistic-level simulations~\cite{Hwang.2017, Zhang.2024}. Such simulations are limited by the model size and simulation time which can render it difficult to observe any phase transitions. Recent studies leveraged machine learned force-field aided \textit{ab initio} molecular dynamics~(MD) simulations at elevated temperatures and observed gliding of layers~\cite{Tang.2024} or employed nudged elastic band~(NEB) calculations to estimate the phase-transition barrier~\cite{Langella.2025}. However, detailed insights about the sodium-concentration dependency on the transition barrier and the transition mechanism at dynamic conditions are still missing due to the associated computational costs. Even though machine learning potentials gained significant interest over the past years and allow for significant speed-ups over quantum-mechanic calculations while often maintain reasonable accuracy, most machine learning potentials are still significantly slower than simple classical potentials~\cite{Fellman.2025, Li.2025, Deringer.2019, Behler.2016, Mueller.2020}. It must be mentioned that in spite of the faster simulations using classical potentials often a lower accuracy than in machine learning potentials is achieved and classical potentials are less transferable and require notorious fitting for new materials. 
Nevertheless, a few attempts have so far been made to describe (layered) oxide cathode materials with the help of classical potentials~\cite{Lee.2012, Kerisit.2014, He.2019, Hart.1998, Morgan.2022}. More recently, also some works focused on classical pairwise potentials to investigate sodium layered oxide cathodes~\cite{Sau.2022, Masese.2021, Sau.2022b}. Despite several challenges these classical potentials offer the opportunity to study cathode materials at atomistic level and at timescales and system sizes that cannot be achieved yet by other methods. To the best of our knowledge, such classical potentials have not been applied to study phase transitions in layered oxide sodium-cathodes yet but bear the potential for more direct atomistic insights under more realistic time and temperature conditions. 

In this work, we developed and applied a classical pairwise potential to simulate and understand the O2-P2 phase transition in the prototype sodium-ion cathode-material Na\textsubscript{\textit{x}}CoO\textsubscript{2}. The Co-based layered oxide is among the most studied simple cathode materials with many of literature reports available and is therefore a suitable material for this study~\cite{Delmas.1981, Boddu.2021, Kaufman.2019, Rivadulla.2003, Toumar.2015}. Moreover, Na\textsubscript{\textit{x}}CoO\textsubscript{2} has less additional complexities as it is not Jahn-Teller active and charges are usually just delocalized over all Co ions (according to higher level \textit{ab initio} calculations~\cite{Koster.2024}). While the material is reported in the P2, O3, and P3 phase, also phase transitions during desodiation are reported~\cite{Delmas.1981, Biecher.2022} making it a suitable candidate to study phase transitions. Even though no full O2 phase (just irregular OP phase at high states of charge~\cite{Biecher.2022}) was proposed for this material yet it still seems reasonable to study the O2-P2 transition as the mechanisms should be similar to other O\textit{n}-P\textit{n} transitions and the O2-P2 case is the most simple to starting point with just one layer that is gliding~\cite{Langella.2025}. Moreover, the O2-P2 phase transition is of great relevance to layered sodium-ion cathode-materials in general~\cite{Zhang.2024, Lee.2013}. To that extend, this study can be seen as a basis to study more complex materials and phase transitions while we focus here on the most simple cases to outline fundamentals of phase transitions in layered oxides and possible methodologies how these could be studied.

We begin with a detailed methodology section describing the fitting procedure of a Coulomb-Buckingham potential for the O2-P2 Na\textsubscript{\textit{x}}CoO\textsubscript{2} system. We continue by performing static calculations using density functional theory~(DFT) and by employing the fitted potential to learn about how the phase-transition barrier changes with sodium concentration. Next, the fitted potential is applied in classical MD simulations to gain mechanistic insights and to obtain the activation energy for phase transitions under dynamic conditions. Finally, we show that with the help of classical potentials it is possible to observe phase transition under standard lab conditions at the time scale of $\upmu$s. Moreover, we also show that the fitted potential is capable to reproduce sodium-diffusion properties and captures the different sodium conductivity of O- and P-phases well. The simplified workflow followed in this work is sketched in Figure~\ref{fig:abstract}.

\section{Computational Methods}
\begin{figure*}[tb]
\includegraphics[width=0.9\textwidth]{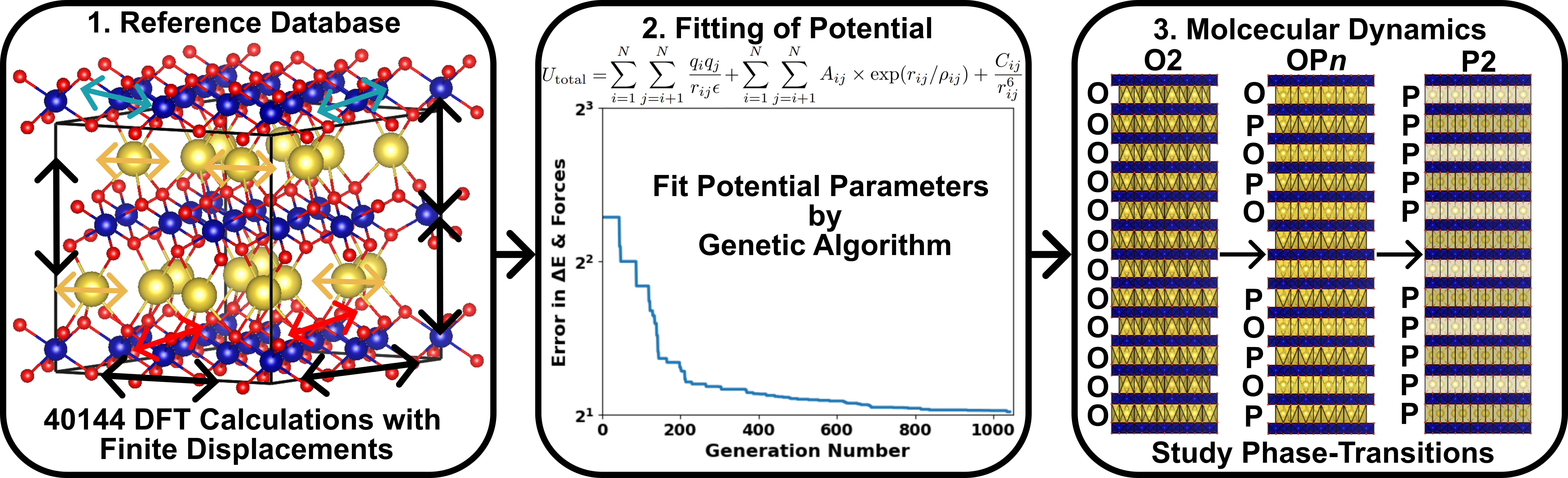}
\caption{\label{fig:abstract} Schematic procedure of database creation, potential fitting, and application in molecular dynamics~(MD) for the Na\textsubscript{\textit{x}}CoO\textsubscript{2} cathode material performed in this work. First, more than 40000 structures with finite displacements of lattice parameters and ion positions (indicated by arrows in the figure) were evaluated by density function theory~(DFT) calculations. Second, a Coulomb-Buckingham potential was fitted to energy differences and forces of the reference database leveraging a Genetic Algorithm~(GA). Finally, the obtained potential was applied in large-scale classical MD simulations in the $\upmu$s regime to study phase transitions.}
\end{figure*}

The general computational methodology employed in work is outlined in Figure~\ref{fig:abstract} consisting of three steps: creation of an \textit{ab initio} reference dataset, fitting of the Coulomb-Buckinghma potential, and performing MD simulations. While the first two steps are discussed in some more detail in the following subsections of this section, MD simulations are analysed and discussed in the next section (Results and Discussion).

\subsection{The employed Potential}

The proposed potential follows a conventional Coulomb-Buckingham potential approach for all pairwise interactions in the Na\textsubscript{\textit{x}}CoO\textsubscript{2} cathode material. The potential is thus given by:

\begin{equation}
    U_{\text{total}} = U_{\text{Coulomb}} + U_{\text{Buckingham}}.
\end{equation}

The two different contributions can be further specified by:

\begin{equation}
    U_{\text{Coulomb}} = \sum_{i=1}^{N} \sum_{j=i+1}^{N} \frac{q_iq_j}{r_{ij}\epsilon}
\end{equation}

where the sums ensure to consider all pairwise interactions of the $N$ ions in the system with $q_i$ and $q_j$ being the ionic charges of the corresponding ions, $r_{ij}$ is their euclidean distance and $\epsilon$ is the dielectric constant of the material. For the ionic charges, always the bare formal charge (no effective charge) was applied and $\epsilon$ was fitted as a free parameter. Fixed charges of $1+$ for Na and $2-$ for O were assumed and a variable charge between $3+$ and $4+$ for Co to charge-balance the total structures at all sodium concentrations. To account for the long-range character of electrostatic interactions and the periodic boundary conditions Ewald summation~\cite{Ewald.1921} is applied to solve the Coulomb potential contribution. As Coulomb interactions missing the Born repulsion at very short distances, Buckingham potentials are employed that have the following form~\cite{Buckingham.1938}:

\begin{equation}
    U_{\text{Buckingham}} = \sum_{i=1}^{N} \sum_{j=i+1}^{N} A_{ij} \times \text{exp}(-r_{ij}/\rho_{ij}) - \frac{C_{ij}}{r_{ij}^6}.
\end{equation}

In the Buckingham potential again $r_{ij}$ denotes the euclidean distance of the two ions $i$ and $j$ that are currently considered in the double-sum over all pairwise interactions. $A_{ij}$, $\rho_{ij}$, and $C_{ij}$ are free parameters that need to be fitted for all pairs of different ionic species in the system. In this work, parameters are fitted explicitly for each pair of different species and no interpolation is applied.

\subsection{Creation of Reference Dataset}

Considering the described Coulomb-Buckingham potential approach, reference data for fitting the potential parameters in the Na\textsubscript{\textit{x}}CoO\textsubscript{2} layered oxide cathode-system is required. To capture the most important features of the material, namely different sodium concentration, sodium orderings, sodium diffusion, phase transitions, and lattice parameter changes a broad variety of reference structures is required. Hence, structures in O2, P2, and OP2 stacking phases were modelled. Moreover, five equidistant intermediate structures between O2 and P2 were modelled (e.g., different shifts of the two transition-metal layers towards each other) to account for dynamic phase transitions. 

The structures were represented by $2\sqrt{3}\times2\sqrt{3}\times1$-supercells containing 24 formula units of Na\textsubscript{\textit{x}}CoO\textsubscript{2}. For all of these eight phases different sodium concentrations of \textit{x} = 1.00, 0.83, 0.67, 0.50, 0.33, 0.17, and, 0.00 in Na\textsubscript{\textit{x}}CoO\textsubscript{2} were included in the reference dataset. To account for different sodium orderings up to ten (e.g., there are no sodium orderings in \textit{x}=0.00) local minimum sodium orderings were considered for each material. The different sodium arrangements were determined by optimizing the electrostatic energies with help of the gradient decent algorithm implemented in the GOAC (Global Optimization of Atomistic Configurations by Coulomb) code~\cite{Koster.2025} while only local minima orderings that differ by at least 0.1~eV in electrostatic energy were chosen. This procedure resulted in a total of 386 reference structures in the Na\textsubscript{\textit{x}}CoO\textsubscript{2} layered oxide cathode-system. Finally, the reference dataset was created by generating randomly distorted samples for all of the 386 structures. 
In order to do so, 10~\% (25~\% for strong distortions) of the ions were selected and displaced randomly by up to 0.25~\AA. To capture sodium migration and the sodium diffusivity, 20~\% (75~\% for strong distortions) of the sodium ions were selected randomly and displaced by up to 0.5~\AA (0.75~\AA for strong distortions) with a two times (4.5 times for strong distortions) higher magnitude for displacements along the \textit{a}- and \textit{b}-direction than along the \textit{c}-direction. Moreover, the lattice parameters were distorted randomly as well in the sample structures by $\pm$2~\% ($\pm$5~\% for strong distortions) in the \textit{a}- and \textit{b}-direction. Along the \textit{c}-direction inter- (O-Na-O) and intra-layer (O-Co-O) separations were varied for all layers each by up to $\pm$3~\% ($\pm$7.5~\% for strong distortions). These inter- and intra-layer distortions were employed as they are expected to undergo the most pronounced changes during phase transitions and desodiation. For the strong distortions also lattice angles were varied by $\pm$2.5~\%. For the initial structures the distortions were applied to, DFT optimized lattice parameters of \textit{a}=\textit{b}=4.875~\AA, \textit{d}\textsubscript{O-Co-O}=1.95~\AA, and \textit{d}\textsubscript{O-Na-O}=3.45~\AA~ long with perfectly symmetric ion positions were considered. These lattice parameters were calculated as the average values of O2 and P2 stackings at \textit{x} = 0.50 as obtained from DFT optimizations. To create the reference dataset for fitting the potential parameters 104 (52 with normal and 52 with strong distortions) randomly distorted samples were created for each of the 386 structures resulting in a total of 40144 reference calculations. The methodology of these distortions of ionic positions and lattice parameters is also indicated by the arrows in the first step shown in Figure~\ref{fig:abstract}. All sample structures were evaluated by spin-polarized DFT single-point calculations employing the projector-augmented wave (PAW) method~\cite{Blochl.1994} as implemented in the Vienna \textit{ab initio} simulation package (VASP)~\cite{Kresse.1996}. The Perdew-Burke-Ernzerhof~(PBE) exchange-correlation functional~\cite{Perdew.1996} was applied along with an energy-cutoff of 600~eV and an electronic convergence criterion of $10^{-6}$~eV. The $\Gamma$-Point and standard pseudo potentials were employed. In addition, the D2 dispersion correction~\cite{Grimme.2006} was applied to capture the long-range layer-layer interactions more precisely and as a similar term appears in the Buckingham potential.  

\subsection{Error Metric}
Using the reference dataset (cf. Figure~\ref{fig:abstract}, first step), we fitted the Coulomb-Buckingham potential parameters by minimizing the root‑mean‑square error~(RMSE) between DFT‑computed energy differences (for sample structures at a specific sodium concentration) and each element of the force matrix. Energy differences between samples at different sodium concentrations were excluded as the described potential does not include any term that could account for the ionization energy required to oxidize cobalt. Thus, the scope of the potential is to accurately describe the dynamics, e.g., sodium diffusion and phase transitions, at a given sodium concentration. For the RMSE calculation, the RMSE in forces was weighted ten times more as the elements in the force matrix are usually smaller than energy differences and forces are also rather important to obtain correct dynamics when applying the fitted potential. In total, the error metric that was minimized during fitting the potential parameters can be described by:

\begin{eqnarray*}
    \text{Error} &=& \sum_{x \in X} \frac{S_x}{S} \times\\
     & &\sigma \times \sqrt{ \frac{\sum_{s=1}^{S_x} \sum_{i=1}^{N_{x}} \sum_{d=1}^{3} (f_{xsid}-f_{xsid}^{\text{DFT}})^2}{S_x \times N_{x} \times 3}} + \\
    & &\sqrt{\frac{\sum_{s_1=1}^{S_x}\sum_{s_2=s_1+1}^{S_x} (\Delta E_{xs_1s_2}-\Delta E_{xs_1s_2}^{DFT})^2}{\sum_{s_1=1}^{S_x}\sum_{s_2=s_1+1}^{S_x} N_{x}}}.
    \label{eq:one}
\end{eqnarray*}

In this Equation, $x$ denotes the sodium concentration and the following sodium concentrations were included: $x=$\{0.00, 0.17, 0.33, 0.50, 0.67, 0.83, 1.00\}. $S$ is the total number of samples~(40144) and $S_x$ is the number of samples at a given sodium concentration $x$. The first squared term describes the RMSE of the force matrix $f$ relative to the DFT reference force matrix $f^{DFT}$ weighted by a factor $\sigma=10$. For the given sodium concentration $x$, all samples $s$, all ions $i$, and all three dimensions $d$ are considered in the RMSE of the forces. The second squared term accounts for the RMSE in energy differences between the samples $s_1$ and $s_2$ at the given sodium concentration $x$. The RMSE is normalized to per ion energies by dividing by the total number of ions in the system ($N_x$). The exact formulation of the error metric is somewhat arbitrary but proved to be efficient in fitting the potential parameters.

\subsection{Fitting of Potential Parameters}

\begin{figure*}[tb]
\includegraphics[width=0.975\textwidth]{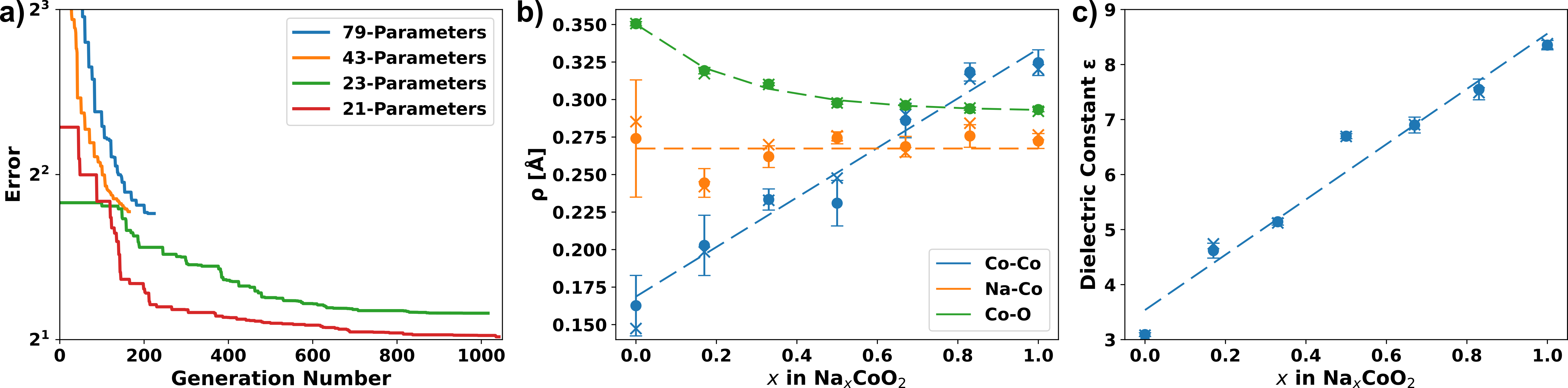}
\caption{\label{fig:fitting} Fitting plots of the process of obtaining the potential parameters. a) Convergence plots of Genetic Algorithm for fitting different potential models with decreasing numbers of parameters. b) Dependence of $\rho$ in interactions with Co on sodium concentration as obtained from the potential model with 43 parameters. Crosses indicate the parameter value of the best fit, circles the mean value of the first 1000 best optimized parameter sets and error bars their standard deviation. Fitted functions that were considered in further potential models are shown by dashed lines. c) Dependence of $\epsilon$ on sodium-concentration as obtained from the potential model with 43 parameters. Symbols, error bars and lines correspond to same quantities as in plot b).}
\end{figure*}

With the potential form, the \textit{ab initio} reference dataset, and the error metric described in the previous subsections, fitting of the potential parameters can be attempted (second step in Figure~\ref{fig:abstract}). As mentioned before, classic ionic charges~($q$) and no effective charges that would require fitting were considered. Still, this approach yields 19~($1+3\times6$) potential parameters, e.g., $\epsilon$ and 3 Buckingham parameters for 6 pairs (Na-Na, Na-Co, Na-O, Co-Co, Co-O, O-O). Moreover, a different set of parameters is required for each sodium concentration in Na\textsubscript{\textit{x}}CoO\textsubscript{2}. Thus, fitting the potentials for all of the 7 considered sodium concentrations included in the reference dataset would require fitting of 133~($7\times$19) parameters. However, as only $q_{\mathrm{Co}}$ is changing with different sodium concentrations, e.g., Co is oxidized and reduced between Co\textsuperscript{3$+$} and Co\textsuperscript{4$+$}, it seems justified to assume that Buckingham parameters for pairs that do not include Co are the same among all sodium concentrations. This greatly reduces the number of potential parameters for Na\textsubscript{\textit{x}}CoO\textsubscript{2} from 133 to a total of 79 parameters. However, this assumption also requires a coupled fitting of all sodium concentrations at the same time as certain optimized parameters must be the same for all sodium concentrations. It was attempted to fit this potential by global optimization using the genetic algorithm (differential evolution) as implemented in the scipy package\cite{Virtanen.2020, Storn.1997}. A population size of 512 samples (256 samples for 23 and 21 parameter models) was used along with "sobol" initialization. For mutations dithering was employed in the range of 0.25 to 1.5 along with a recombination constant of 0.6. The best sample of the population was optimized by the L-BFGS (limited memory Broyden–Fletcher–Goldfarb–Shanno) algorithm~\cite{Liu.1989} after each iteration. The following boundaries for Buckingham parameters of $A$, $\rho$, $C$, and $\epsilon$ were applied: [100, 30000], [0.1, 0.5], [0, 1000], and [1, 10], respectively. The Buckingham cut-off was set to 15~\AA while electrostatics part was solved exactly by Ewald summation. Convergence of the fitting involving 79 parameters can be seen in Figure~\ref{fig:fitting}~(a).

As 79 parameters describing the Na\textsubscript{\textit{x}}CoO\textsubscript{2} system are still quite inconvenient, it was tested if it is possible to fit a potential assuming that also the $C$ Buckingham-parameters for pairs including Co are the same for all sodium concentrations. This assumption might be justified by the fact that also in the D2 dispersion correction the term $C$/$r^6$ is used while $C$ is independent of the chemical environment, e.g., the ionic charge~\cite{Grimme.2006}. Fixing the $A$ parameters for pairs including Co as well does not significantly reduce the degrees of freedom in the fit as this fix can be mostly compensated by adjusting the $\rho$ parameters. Ultimately, these assumptions result in a potential with 43 parameters and convergence of the error metric for fitting this model is shown in Figure~\ref{fig:fitting}~(a) as well. Even though both fittings, the one of the 79 parameter model and the one of the 43 parameter model, were stopped early, their similar convergence behaviour shown in Figure~\ref{fig:fitting}~(a) indicates that the model with 43 parameters still has enough degrees of freedom to capture the characteristics of the Na\textsubscript{\textit{x}}CoO\textsubscript{2} system.

We further tried to reduce the parameters of the Na\textsubscript{\textit{x}}CoO\textsubscript{2} potential by analysing the variation of the remaining sodium concentration dependent parameters ($\rho_{\text{Co-Co}}$, $\rho_{\text{Na-Co}}$, $\rho_{\text{Co-O}}$, $\epsilon$) as function of sodium concentration $x$. The $\rho$ parameters for all pairs involving Co are plotted in Figure~\ref{fig:fitting}~(b) and indicate that $\rho_{\text{Na-Co}}$ is sufficiently described by a fixed value over the whole sodium range reducing the 7 parameters for each sodium concentration to just 1. Moreover, $\rho_{\text{Co-Co}}$ has the strongest dependency on sodium concentration while its change is well described by a simple linear function of the form $f(x)=ax+b$. Thus, the 7 parameters per sodium concentration can be reduced to two parameters that describe the linear function, namely $a$ and $b$. Lastly, $\rho_{\text{Co-O}}$ shows a small change on varying the sodium concentration that is apparently well described by an exponential decay of the general form $f(x)=a\exp(bx)+c$, transforming the 7 fitted parameters per sodium concentration to the 3 parameters of the function: $a$, $b$, $c$. Finally, the fitted dielectric constants undergo strong changes at different sodium concentrations which are visualized in Figure~\ref{fig:fitting}~(c). The trend of $\epsilon$ and sodium concentration $x$ is almost perfectly linear indicting that the 7 $\epsilon$ for each sodium concentration in the fitting can be replaced by just 2 parameters that describe the linear function, similar to $\rho_{\text{Co-Co}}$. The approach of fitting functions instead of parameters for each sodium concentration bears two substantial advantages: I) The total number of parameters in fitting is reduced which speed-ups convergence and II) the resulting potential is more generalized as the functions allow to interpolate to intermediate sodium concentrations that were not considered in the reference dataset used for the fitting.

\begin{table*}[tb]
    \centering
    \caption{Coulomb-Buckingham potential for Na\textsubscript{\textit{x}}CoO\textsubscript{2} fitted and employed in this work.}
    \setlength{\tabcolsep}{12pt}
    \begin{tabular}{c|ccc}
    \toprule
        Pair & $A$ & $\rho$ & $C$ \\
    \hline
        Na-Na & 101.02 & 0.33982 & 0.31577 \\
        Na-Co & 701.536 & 0.32767 & 6.6555 \\
        Na-O & 6276.7 & 0.22321 & 40.010 \\
        Co-Co & 8582.7 & 0.019341$x+$0.27366 & 403.57 \\
        Co-O & 11474 & 0.19886 & 41.395 \\
        O-O & 4144.2 & 0.23746 & 4.1772 \\
    \toprule
    Charge & Na & Co & O \\
    $q$ & $+$1 & +4$-x$ & $-$2 \\
    \hline
    & & $\epsilon$=1.0379$x$+8.1337 & \\
    \toprule
    \end{tabular}
    \label{tab:Potential}
\end{table*}

Considering these functions and fixes for the parameters the total number of parameters in the potential fitting reduces further from the aforementioned 43 parameters to 23 parameters and the corresponding convergence curve is shown in Figure~\ref{fig:fitting}~(a). The pronounced and fast convergence towards errors significantly lower than the models with more parameters prove that the assumptions made for the parameter functions do not decrease the flexibility of the potential too much and still allow for a reasonable fitting of the potential. Ultimately, the exponential function assumed for $\rho_{\text{Co-O}}$ became very flat during fitting indicating that $\rho_{\text{Co-O}}$ might be considered a fixed parameter over the whole sodium range as well, reducing the total amount of parameters of the potential to just 21. Indeed, the convergence plot in Figure~\ref{fig:fitting}~(a) highlights that also with a fixed value of $\rho_{\text{Co-O}}$ a potential with a low error can be fitted, even significantly lower than for the model with 23 parameters. This might be explained by the fact that an exponential function is more difficult to fit and very sensitive to small parameter changes. The fitting without the exponential function for $\rho_{\text{Co-O}}$ shows therefore a better convergence behaviour and an overall lower error metric for the final potential. The resulting Coulomb-Buckingham-Potential parameters which have been applied in this work are listed in Table~\ref{tab:Potential}. Notably, also the change of $\rho_{\text{Co-Co}}$ with the sodium-concentration is small, indicating the possibility of fitting Coulomb-Buckingham potentials for different sodium concentrations of layered oxides by just varying the charges of the transition metals as well the dielectric constant in a linear fashion. However, in this work we also use a linear function for $\rho_{\text{Co-Co}}$ as the resulting potential appears to be well-fitted (cf. Fig.~\ref{fig:fitting}~(a)) and sufficiently simplified. The fitted average (over sodium concentration) effective dielectric constant of approx. 8.65 is in very good agreement to previously reported first principles studies~\cite{Li.2004} that suggest an average (over structure models and lattice directions) dielectric constant of 8.86 while other computational studies~\cite{Petousis.2017} also reported lower values of 5.66 (for octahedral sodium coordination).

\begin{table}[tb]
    \centering
    \caption{Coulomb-Buckingham potential with a Buckingham potential for Na-O obtained from fitting a Coulomb-Buckingham potential to Na\textsubscript{2}O. The described potential is also applied to Na\textsubscript{\textit{x}}CoO\textsubscript{2} in this work.}
    \setlength{\tabcolsep}{12pt}
    \begin{tabular}{c|ccc}
    \toprule
        Pair & $A$ & $\rho$ & $C$ \\
    \hline
        Na-O & 2566.3 & 0.27302 & 67.186 \\
    \toprule
    Charge & Na & Co & O \\
    $q$ & $+$1 & +4$-x$ & $-$2 \\
    \hline
    & & $\epsilon$= 2.2733 & \\
    \toprule
    \end{tabular}
    \label{tab:Potential-Na2O}
\end{table}

To determine which pairwise interactions are most important and to understand how much complexity of a potential is actually required to study phase transitions in layered oxides we fitted another Coulomb-Buckingham potential to 100 distorted structures of Na\textsubscript{2}O. While a full potential with parameters for all pairs (Na-Na, Na-O, and O-O) was fitted, only the Buckingham potential for the Na-O interaction was considered in the tests and the resulting total potential is described in Table~\ref{tab:Potential-Na2O}. Again, the fitted dielectric constant of 2.27 is in comparable order of magnitude to previously reported calculations that determined a value of 3.27~\cite{Petousis.2017}. This potential can be seen as the most simplistic potential that allows to study phase-transition barriers by only adding one Buckingham potential to bare electrostatic interactions and by assuming that Na-O interactions are mostly independent on structure-type and coordination. Thus, the potential has also no dependencies on the sodium concentration in Na\textsubscript{\textit{x}}CoO\textsubscript{2}.

\subsection{Calculation of Phase-Transition Barriers}

Calculation of static phase-transition barriers was achieved by starting with an O2 structure and shifting one CoO\textsubscript{2} layer as a whole by \textit{a}/3 along the \textit{a}-axis which leads a P2 structure. A total of 21 equidistant images along this reaction path were considered in $2\sqrt{3}\times2\sqrt{3}\times1$-supercells containing 24 formula units of Na\textsubscript{\textit{x}}CoO\textsubscript{2}. At each image, sodium-ordering was determined over both sodium sub-lattices of the P2 structure while one sub-lattice coincides with the one of O2 which was fixed and the other was shifted proportional to the reaction coordinate. Sodium orderings were optimized by configurational optimization of the total Coulomb energy assuming the charges described in Table~\ref{tab:Potential}. Replica Exchange Monte Carlo (REMC) as implemented in the GOAC package~\cite{Koster.2025} with parallel tempering at 0.1, 0.2, 0.3, 0.4, 0.6, 0.8, 1.0, and 1.2~eV was leveraged for the configurational optimization. Obtained lowest energy configurations for each image were than evaluated by DFT with the same settings described before, the Coulomb-Buckingham potential, and the potential considering just the Na-O Buckingham interaction and electrostatics. For comparison, the lattices of the cells (just cell volume) at each image were relaxed by DFT with an convergence criterion of a $10^{-5}$~eV change in total energy. Furthermore, the lattices were also relaxed by the fitted Coulomb-Buckingham potential with help of the LAMMPS software package~\cite{Thompson.2022, Plimpton.1995}. A Buckingham cutoff of 15~\AA along with Ewald summation for electrostatic interactions were applied. The fractional volume change was limited to $10^{-3}$ for each iteration of the FIRE algorithm~\cite{Bitzek.2006}.

\subsection{Molecular Dynamics Simulations}

For the MD simulations $4\sqrt{3}\times4\sqrt{3}\times6$-supercells of Na\textsubscript{0.67}CoO\textsubscript{2} with 576 formula units (2112 atoms) were created. Simulations were carried out in the LAMMPS software package~\cite{Thompson.2022, Plimpton.1995} as well with the same settings described for the minimizations in the subsection before. For each simulation the structure was first fully minimized (lattice and ionic positions), followed by an MD relaxation of 30~ps during which the temperature was linearly increased from 0~K to the target temperature. The relaxations were performed in the \textit{npT} ensemble with an isotropic pressure of 1~bar. Both, temperature and pressure were controlled by a Nose-Hover-thermostat (barostat)~\cite{Nose.1984, Hoover.1985} with damping parameters of 0.1~ps for the temperature and 1~ps for the pressure. LAMMPS applies the equations of motions as described by Shinoda~\textit{et al.}~\cite{Shinoda.2004} along with the time integration schemes described by Tuckerman~\textit{et al.}~\cite{Tuckerman.2006} while a time step of 2~fs was used in this work. After the relaxation runs MD production runs were carried out at the fixed target temperature and a pressure of 1~bar in the \textit{npT} ensemble and by the same settings described for the relaxation runs. Snapshots of the simulation were written every 1~ps and structures shown in this work were visualized with help of the VESTA software~\cite{Momma.2008}.

\section{Results and Discussion}

\subsection{Calculation of Phase-Transition Barriers by Static Simulations}

\begin{figure*}[tb]
\includegraphics[width=0.975\textwidth]{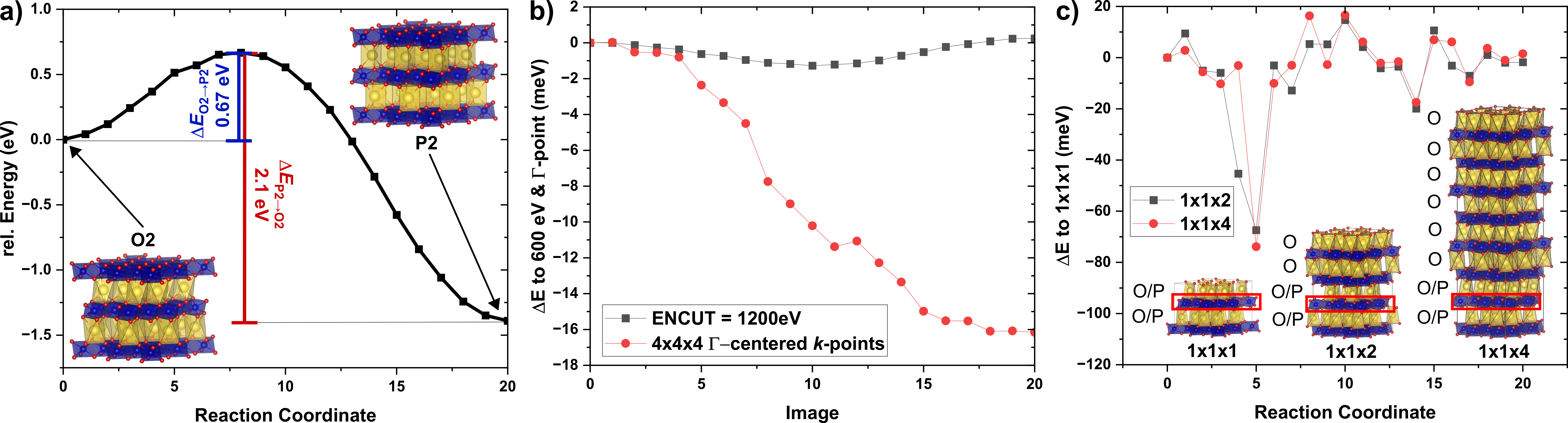}
\caption{\label{fig:Convergence} Visualization of a reaction pathway and computational convergence considerations. a) Reaction path of the O2-P2 phase transition as obtained from DFT calculations of the perfect images (no relaxation of lattice parameters) normalized to the energy of O2 at a sodium concentration of 0.67. The O2 to P2 ($\Delta E_{\text{O2} \to \text{P2}}$) as well as the P2 to O2 ($\Delta E_{\text{O2} \to \text{P2}}$) phase-transition barriers can be obtained from the reaction path and are indicated by blue and red lines, respectively. b) Convergence of the applied DFT settings with respect to energy cut-off and \textit{k}-point sampling. Energy differences to the plot in a) are shown. c) Convergence with respect to supercell size in \textit{c}-direction by treating more O-layers beyond the transitioning layer explicitly. The transitioning layer is highlighted by a red box in the shown structure models.}
\end{figure*}

First, we studied the O2-P2 phase-transition reaction-path of Na\textsubscript{\textit{x}}CoO\textsubscript{2} in static cells and an exemplary energy transition-pathway is shown in Figure~\ref{fig:Convergence}~(a) for the O2-P2 transition in Na\textsubscript{0.67}CoO\textsubscript{2} evaluated by DFT single-point calculations. The resulting reaction pathway shows the expected thermodynamic stabilities between the O2 and P2 phase with the P2 phase being almost 1.5~eV more stable. This matches experimental studies that also report the P2 phase to be stable at intermediate sodium concentrations such as 0.67~\cite{Biecher.2022, Berthelot.2011}. However, the reaction path also reveals a barrier for the transition of the O2 to P2 phase while the pure O2 phase is a local minimum that is kinetically stabilized through the aforementioned barrier. Thus, the reaction path allows to determine the transition barrier ($\Delta E$) between the two (local) minima in both directions (O2 to P2 and P2 to O2) as indicated by the blue and red lines in Figure~\ref{fig:Convergence}~(a). The corresponding phase transition barriers are in the following referred to as $\Delta E_{\text{O2} \to \text{P2}}$ and $\Delta E_{\text{P2} \to \text{O2}}$, respectively. Due to the higher thermodynamic stability of the P2 phase, $\Delta E_{\text{P2} \to \text{O2}}$ is more than three times larger in this example but also the O2 to P2 transition shows a considerable barrier of $\Delta E_{\text{O2} \to \text{P2}}$=0.67~eV highlighting that despite the much higher stability of the P2 phase the transition is still not barrier-free. The transition state of the phase transition is shifted towards the O2 phase as well because of the higher stability of P2.

To verify that the employed DFT settings are sufficiently converged to allow computation of $\Delta E$ for the O2-P2 phase transition, a convergence test with respect to the energy cutoff and \textit{k}-points was carried out. Results are plotted in Figure~\ref{fig:Convergence}~(b) showing only minor deviations of few meV in the calculated $\Delta E$ compared to the employed settings. Towards higher image numbers the \textit{k}-point convergence slightly worsens as the energies were normalized to the O2 phase (image 0). Still, energy differences are far less than 20~meV and corresponding errors in $\Delta E$ are even lower due to partial error cancellations. Moreover, $\Delta E$ are in the range of several hundreds or even thousands of meV and therefore significantly larger than the estimated computational error. Thus, the employed cutoff and \textit{k}-points seem to be sufficiently converged to reliably conclude the trends in $\Delta E$, most likely also supported by error cancellation when normalizing the images to the O2 phase.

Finally, from a practical point of view it is also important to check whether the long-range interactions between the layers along the \textit{c}-direction influence the obtained $\Delta E$ or not. Under realistic conditions the phase transition occurs most likely gradually such that the O2 to P2 transition happens under periodic interaction to further O-layers. In a small cell $\Delta E$ might be affected be the long-range self-interactions of the gliding layer. Accordingly, we checked the convergence with respect to supercell-size along the \textit{c}-direction by treating more and more layers explicitly that remain in the O-phase during the transition of one layer. The results and structural models for supercells that are two- and four-times larger along the \textit{c}-direction are presented in Figure~\ref{fig:Convergence}~(c). Only a small fluctuation (below 20~meV) for all images is found which is negligible in comparison to the absolute values of energy barriers ($\Delta E_{\text{O2} \to \text{P2}}$=0.67~eV or $\Delta E_{\text{P2} \to \text{O2}}$=2.1~eV in the example in Figure~\ref{fig:Convergence}~(a)). However, for the sixth image (reaction coordinate 5) stronger deviations of almost 80~meV are observed. This deviation can be explained by insufficient convergence of the sodium arrangements during electrostatic optimization or of the electronic structure during DFT calculation (or both) in the original data as also a small deviation (irregular increase) in the relative energy in Figure~\ref{fig:Convergence}~(a) can be seen. However, Figure~\ref{fig:Convergence}~(a) also shows that such small deviations usually do not affect the obtained $\Delta E$ much as long as they are not at the minima or transition state. Regarding the influence of the supercell-size along the \textit{c}-direction it can be therefore concluded that $\Delta E$ is not significantly influenced by the neighbouring layers. More generally, this indicates that the phase transition of one layer is mostly unaffected by long-range interactions to other layers. Thus, phase transitions should happen randomly in a layered oxide structure in terms of which layer transitions first/last when many layers are present along the \textit{c}-direction as long as no other factors influence the transition such as inhomogeneous desodiation per sodium layer or defect/surface effects.

\begin{figure*}[tb]
\includegraphics[width=0.975\textwidth]{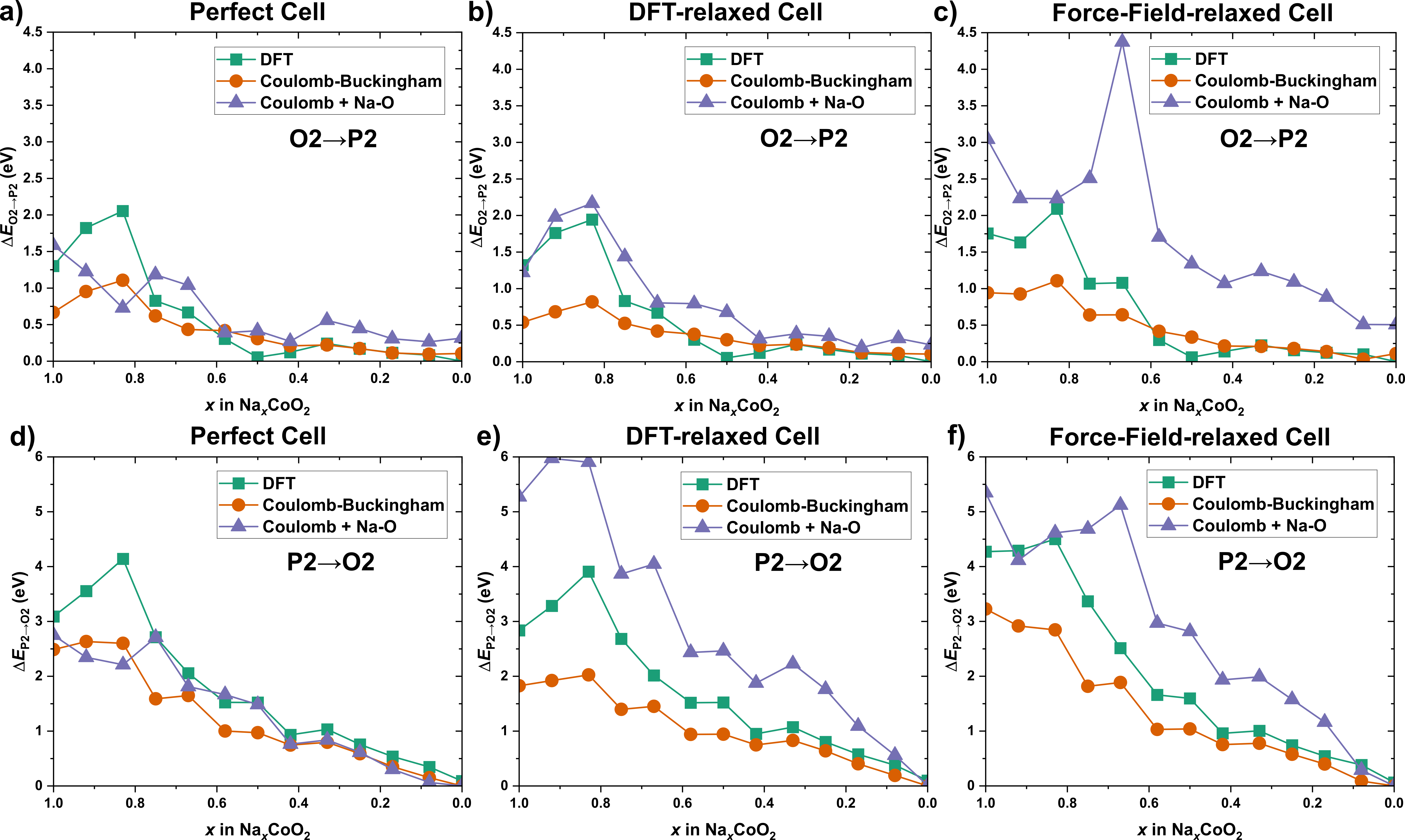}
\caption{\label{fig:barriers} O2-P2 phase-transition barriers in Na\textsubscript{\textit{x}}CoO\textsubscript{2} as function of sodium concentration \textit{x}. Non-relaxed and relaxed cells were considered and all transition barriers were obtained by DFT, the fitted Coulomb-Buckingham potential, and by employing just a Buckingham potential for the Na-O interactions along with full electrostatics. a)-c) O2 to P2 barriers ($\Delta E_{\text{O2} \to \text{P2}}$) and d)-f) P2 to O2 barriers ($\Delta E_{\text{P2} \to \text{O2}}$) for a perfect cell (a) and d)), for a cell with lattice parameters relaxed by DFT for each image (b) and e)), and for cells with lattice parameters optimized with the Coulomb-Buckingham potential fitted in this work (c and f).}
\end{figure*}

Following the methodology sketched in Figure~\ref{fig:Convergence}~(a) to determine $\Delta E$ we continued by calculating $\Delta E$ as function of sodium concentration in Na\textsubscript{\textit{x}}CoO\textsubscript{2} (Figure~\ref{fig:barriers}). An overall trend towards lower $\Delta E$ at higher states of charge (lower sodium concentration) can be observed, which is in agreement with the fact that phase transitions are mostly observed at lower sodium concentrations in experiment~\cite{Zhang.2024}. This indicates that realistic computational predictions of most stable phases should consider both, thermodynamics and kinetics as the larger $\Delta E$ at high sodium concentrations might allow to kinetically stabilize a meta-stable phase under experimental conditions. Moreover, calculated $\Delta E$ are generally higher for P2 to O2 than for O2 to P2 transitions due to the higher thermodynamic stability of P2 over O2 for the Na\textsubscript{\textit{x}}CoO\textsubscript{2} system. Besides this trend towards lower $\Delta E$ with decreasing sodium concentration most graphs in Figure~\ref{fig:barriers} show first an increase of $\Delta E$ from sodium concentrations of 1.00 to 0.83. A possible explanation might be that the P2 phase is thermodynamically less favourable at such high sodium concentrations compared to lower sodium concentrations. This is probably because a more favorable sodium ordering benefiting from occupation of both sodium sub-lattices cannot be realized in the P2 phase due to the strong electrostatic repulsion at the short distances between neighbouring positions of the two different sodium sub-lattices of this phase. Moreover, the O2 to P2 phase transition at a sodium concentration of 0.5 shows consistently lower $\Delta E_{\text{O2} \to \text{P2}}$ in DFT (cf. Figure~\ref{fig:barriers}~(a)-(c)) potentially caused by a distinct electronic structure that is not captured in the force field approaches.

Regarding the accuracy of the potential approaches compared to DFT it can be concluded that the Coulomb-Buckingham potential fitted in this work follows the trends in the DFT $\Delta E$ closely while there is a tendency for under-prediction of $\Delta E$ at high sodium concentrations (sodium concentrations of 0.83 to 1.00). However, especially at low sodium concentrations the fitted Coulomb-Buckingham potential shows promising qualitative and quantitative agreement to DFT results. Interestingly, the approach of using bare electrostatics and adding just a Buckingham potential for the Na-O interaction fitted in the Na\textsubscript{2}O system shows satisfactory performance for predicting $\Delta E$ and its trends (despite a few outliers, e.g., the data point at \textit{x}=0.67 in Figure~\ref{fig:barriers}~(c) or high sodium concentrations in Figure~\ref{fig:barriers}~(e)). Especially at high sodium concentrations predictions are in some cases more close to DFT than the full Coulomb-Buckingham potential, e.g., Figure~\ref{fig:barriers}~(b). This highlights that $\Delta E$ is mainly controlled by the repulsive interaction of sodium and oxygen at short distances (Born repulsion). Sodium and oxygen ions are brought closely together when shifting an entire CoO\textsubscript{2} layer causing the Na-O repulsion to be the determining factor for $\Delta E$. However, as the Na-O interactions should be comparable for different cathode compositions and as only electrostatic interactions for the transition metal~(Co) were considered in this model, this finding implies that kinetic phase stabilization through composition tuning relies on electrostatics (inter-ionic distances and their charges).

The discussed trends in $\Delta E$ are also mostly consistent over different cell geometries especially for the more accurate approaches such as DFT and the full Coulomb-Buckingham potential (c.f. Figure~\ref{fig:barriers}). This proves that the approach of using perfect cells with averaged (over O2 and P2 and at a sodium concentration of 0.5) lattice parameters is generally justified. However, for some sodium concentrations, the absolute $\Delta E$ values change more significantly when relaxation of the lattice parameters is allowed as the initial and final states (O2 and P2) can become thermodynamically more favourable. In order to decrease $\Delta E$ by lowering the energy of the transition state, more strong relaxations than just the lattice parameters are required. This will be discussed in the next section.

\subsection{Calculation of Phase-Transition Barriers by Dynamic Simulations}

\begin{figure*}[tb]
\includegraphics[width=0.975\textwidth]{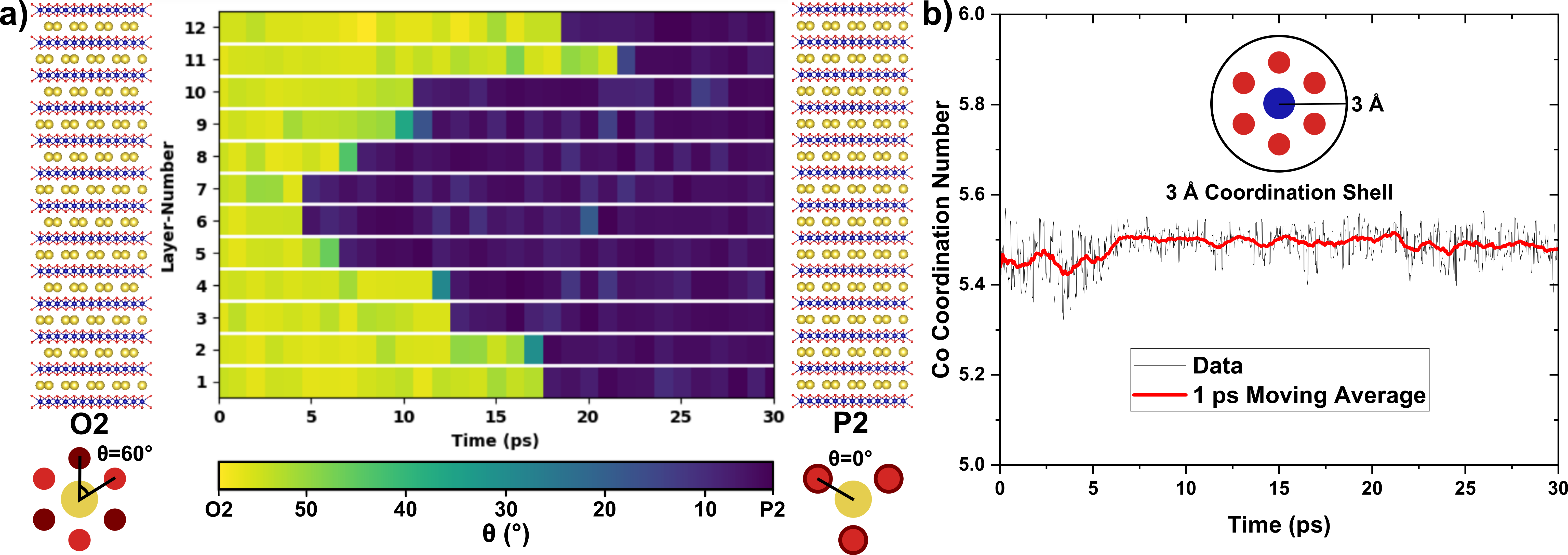}
\caption{\label{fig:Coordination} Sodium and cobalt coordination during a \textit{npT} molecular dynamics simulation at 600~K and 1~bar with fitted Coulomb-Buckingham potential for Na\textsubscript{0.67}CoO\textsubscript{2}. a) Sodium coordination-angles per sodium layer over a runtime of 30~ps. The structure models indicate the sodium layers the bars in the plot correspond to while the sodium coordination is given as the angle $\theta$. As also sketched in the Figure, $\theta$ describes the rotation angle between the upper three (light-red) and lower three (dark-red) oxygen ions that coordinate a sodium ion (yellow). Thus, an angle of 60° corresponds to an octahedral coordination (O2 phase) and an angle of 0° to a prismatic coordination (P2 phase). b) Number of oxygen ions in a coordination sphere of 3~\AA around cobalt ions over run time with the actual data in black and the running average over 1~ps in red.}
\end{figure*}

To allow for pronounced structural relaxations that might occur during the phase transition, we continued by large‑scale MD simulations with 12 layers in the \textit{c}‑direction and using the fitted Coulomb‑Buckingham potential. To obtain results that are as close to experimental conditions as possible \textit{npT} simulations at 1~bar and various temperatures were performed. The fitted Coulomb-Buckingham potential showed promising agreement to DFT in the phase-transition barriers as discussed for Figure~\ref{fig:barriers} but during MD runs at low sodium concentrations the layered structures collapsed after several ps or ns. This most likely highlights that pairwise interactions are insufficient to maintain a 2D layered structure without or little sodium and higher order interactions (3-body and/or 4-body) are required to simulate layered oxide materials at low sodium concentrations. Similar issues were already reported in the literature on delithiation of a layered oxide lithium-ion cathode~\cite{Morgan.2022}. While this finding is important for further theory studies that aim to improve on this issue, the Coulomb-Buckingham potential presented in this work is still able to maintain the atomisitc structure stability during MD simulations at higher sodium concentrations. As there is a strong thermodynamic-driving force (cf. Figure~\ref{fig:Convergence}~(a)) for the O2 to P2 transitions at a sodium concentration of 0.67 and the structures do not collapse at this concentration, in the following MD results for the Na\textsubscript{0.67}CoO\textsubscript{2} material are discussed. Figure~\ref{fig:Coordination}~(a) shows the O2 to P2 phase transitions observed during an MD simulation at 600~K.Firstly, Figure~\ref{fig:Coordination}~(a) proves that with applying our fitted Coulomb-Buckingham potential it is indeed possible to directly observe the O2 to P2 phase transition of Na\textsubscript{0.67}CoO\textsubscript{2} during MD simulations. Moreover, it can be seen that under the applied conditions the transition of a single layer happens quite rapidly (1-3~ps) while the whole structure is gradually transitioning towards P2 \textit{via} different (irregular) types of OP\textit{n} inter-growth phases (often referred to as "Z"-phase~\cite{MortemarddeBoisse.2014}). This consecutive phase-transition mechanism is in agreement with the current gained insight from experimental studies of phase transitions in layered oxide sodium-ion cathodes~\cite{Somerville.2019, Tang.2024}.

The conducted MD simulations allow for more pronounced structural changes during the phase transition and thus might alter the mechanism of gliding an entire CoO\textsubscript{2} layer as assumed for the calculated $\Delta E$ in Figure~\ref{fig:barriers}. Especially the cobalt coordination to nearest neighbor oxygen ions during the transitions is of interest as a recent computational study proposed a transition state with a tetrahedral transition-metal coordination for the P2 to O2 phase transition in Na\textsubscript{\textit{x}}MnO\textsubscript{2}. Therefore, the coordination of Co over simulation time is plotted in Figure~\ref{fig:Coordination}~(b). The coordination, which was approximated by a 3~\AA coordination sphere, is in average slightly below 6 due to artifacts at the cell boundaries as no periodic boundary conditions were considered for the analysis. Most importantly, the coordination number is almost constant over simulation time and always significantly above 5. This proves that during the MD simulations the phase transitions also occurred \textit{via} gliding of an entire CoO\textsubscript{2} layer and no transition states with a tetrahedral cobalt coordination exist. Moreover, the calculated $\Delta E_{\text{P2} \to \text{O2}}$ by DFT at a sodium concentration of 0.17 is 0.54~eV which is significantly smaller than the barrier of 1.04~eV reported for a reaction path involving a tetrahedral transition-metal coordination at a sodium concentration of 0.125~\cite{Langella.2025}. The other reason for the difference in energy barriers might be explained by the fact that a different layered oxide material has been studied by Langella~\textit{et al.}~\cite{Langella.2025}, namely Na\textsubscript{\textit{x}}MnO\textsubscript{2}. In contrast to the studied compound (Na\textsubscript{\textit{x}}CoO\textsubscript{2}), a strong Jahn-Teller activity is expected in Na\textsubscript{\textit{x}}MnO\textsubscript{2}. Further research must show if different transition metals indeed change the phase-transition mechanism or if gliding of an entire TMO\textsubscript{2} layer is the more likely mechanism as indicated by the significantly lower $\Delta E$.

\begin{figure}[tb]
\includegraphics[width=0.475\textwidth]{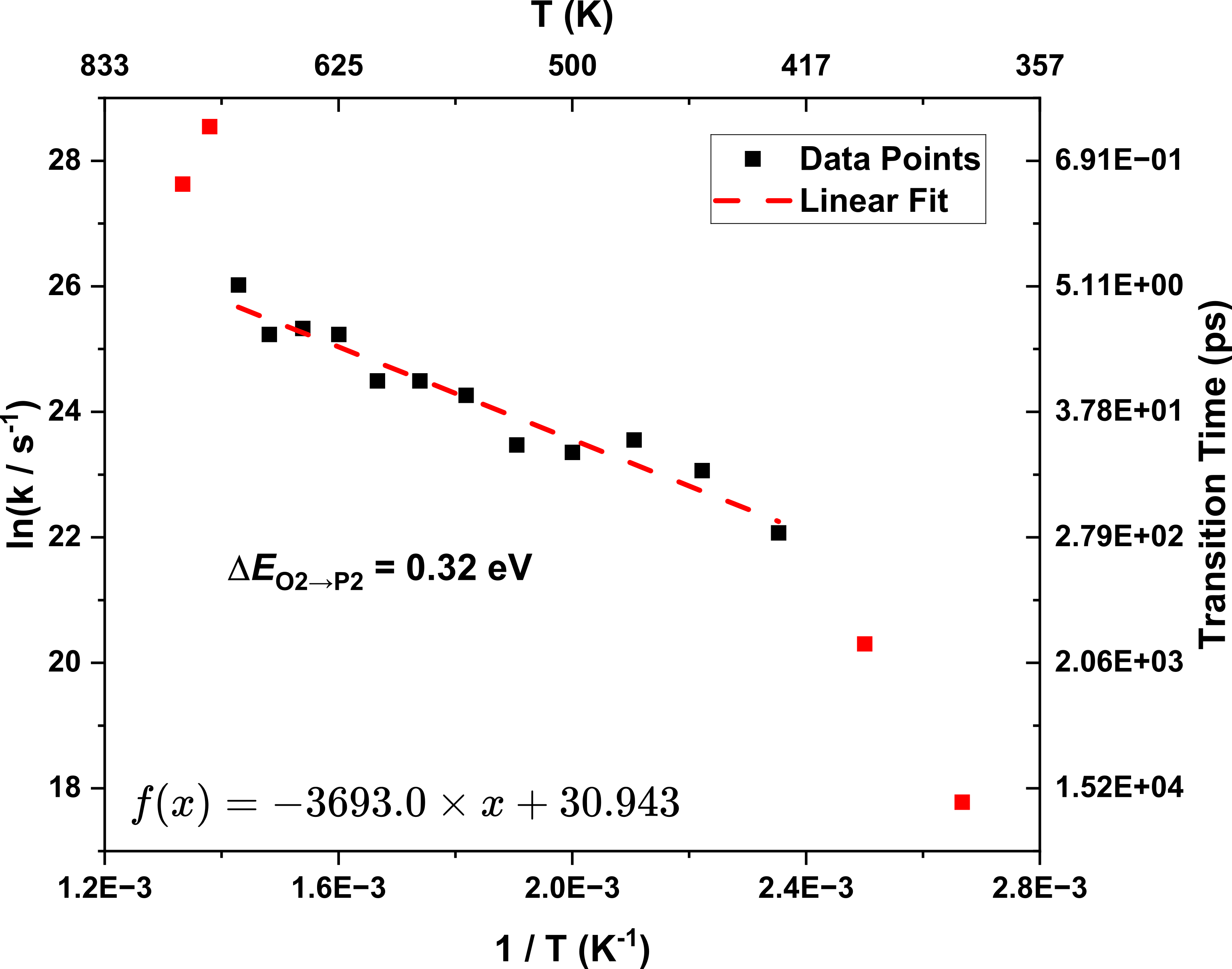}
\caption{\label{fig:Arrehnius} Arrhenius plot for the O2 to P2 phase transition in Na\textsubscript{0.67}CoO\textsubscript{2} obtained from molecular dynamics simulations employing the fitted Coulomb-Buckingham potential. Excluded points are shown in red along with the fitted linear function (dashed red line) and the resulting phase-transition barrier ($\Delta E_{\text{O2} \to \text{P2}}$).}
\end{figure}

To asses whether the increased flexibility of the system during the MD simulations lowers $\Delta E_{\text{O2} \to \text{P2}}$ even further compared to the more static calculations in Figure~\ref{fig:barriers}, we continued by computing an Arrhenius plot for the O2 to P2 phase transition in Na\textsubscript{0.67}CoO\textsubscript{2}. In order to do so, MD simulations in the temperature range of 375~K to 750~K in steps of 25~K were performed and the time to complete the phase transition of the whole structure (e.g., all layers have become prismatic in Figure~\ref{fig:Coordination}~(a)) was tracked to obtain the rate constants ($k$). The corresponding Arrhenius plot is shown in Figure~\ref{fig:Arrehnius}. It can be seen that for the very low temperatures (375~K and 400~K) the transition becomes too sluggish rendering the observed rate constants unreliable while at high temperatures (725~K and 750~K) the transition becomes so fast that it is also difficult to determine accurate timings for the transition. However, in the intermediate temperature range from 425~K to 700~K a clear linear trend can be observed in the Arrhenius plot (cf. Figure~\ref{fig:Arrehnius}). Under the assumption that the phase transition is a first order reaction, which might be justified by the fact that $\Delta E_{\text{O2} \to \text{P2}}$ of the transition of one layer was shown to be mostly unaffected by the other layers in Figure~\ref{fig:Convergence}~(c), the activation energy ($E_{\text{a}}$) can be calculated from a linear fit for the temperature region of 425~K to 700~K by multiplying the obtained slope value by the Boltzmann's constant ($k_{\text{B}}$) following the Arrhenius equation: 

\begin{equation}
    \text{ln}(k) = \text{ln}(A) - \frac{E_{\text{a}}}{k_{\text{B}}T}.
\end{equation}

The calculated $\Delta E_{\text{O2} \to \text{P2}}$ from MD simulation is 0.32~eV, which is lower than those determined by the Coulomb-Buckingham potential in Figure~\ref{fig:barriers}~a)-c): 0.43~eV (perfect cell), 0.42~eV (DFT-relaxed cell), and 0.64~eV (force-field relaxed cell). It is also to expect that the full structural flexibility during the MD simulations allows for lower $\Delta E_{\text{O2} \to \text{P2}}$ than the static and just-optimized the lattice parameters calculations. 
This difference in $\Delta E_{\text{O2} \to \text{P2}}$ shows that additional degrees of freedom in the MD simulations can lower $\Delta E_{\text{O2} \to \text{P2}}$ by more than 25~\% or more than 0.1~eV. An exact quantification of how much dynamic conditions decrease $\Delta E_{\text{O2} \to \text{P2}}$ in general remains difficult. Still, it becomes evident that the values presented in Figure~\ref{fig:barriers} can be understood as an upper limit, while under experimental conditions most-likely even lower $\Delta E$ can be achieved due to additional degrees of freedom in the structural relaxation during the transition.

\begin{figure*}[tb]
\includegraphics[width=0.975\textwidth]{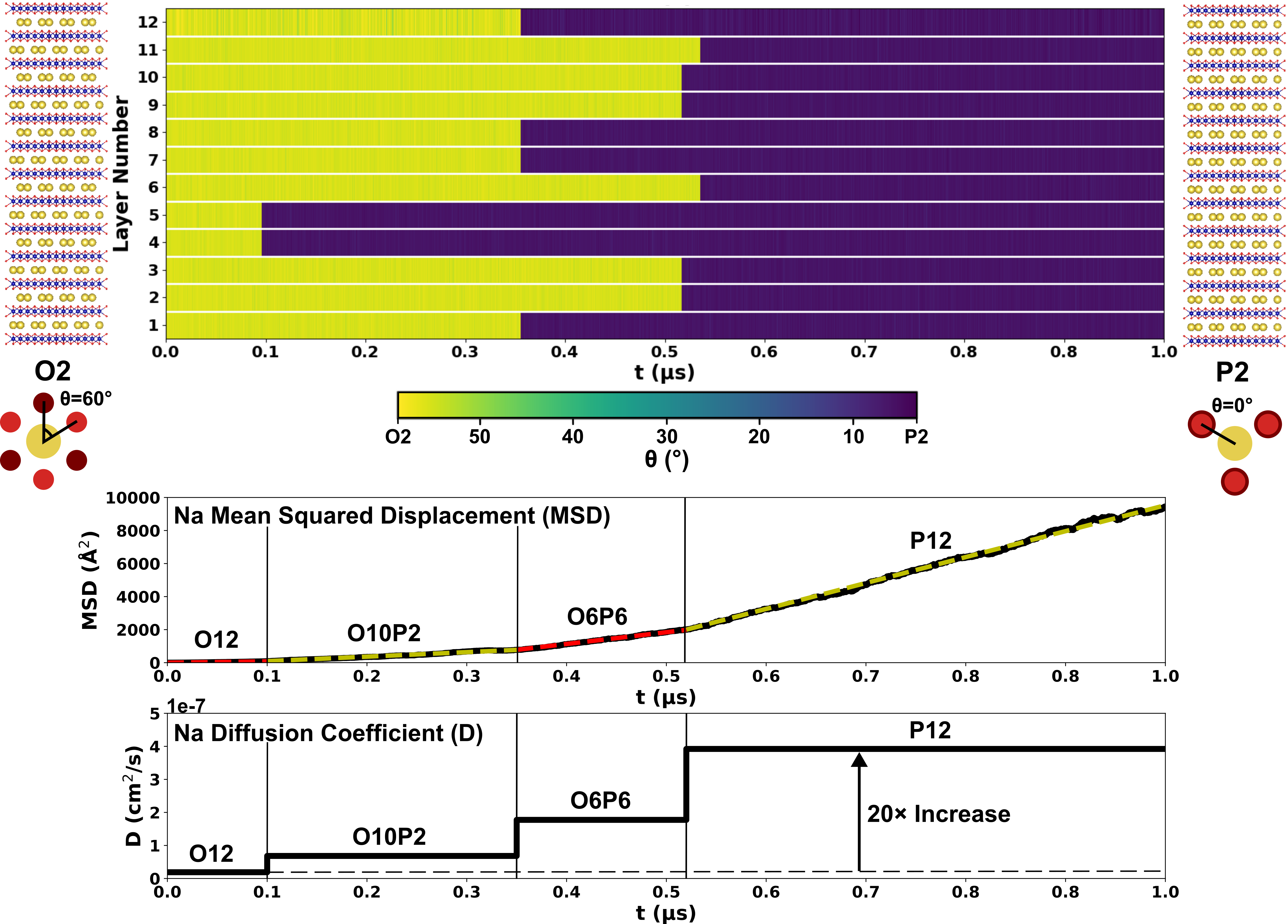}
\caption{\label{fig:MD} O2 to P2 phase transition and sodium diffusion in Na\textsubscript{0.67}CoO\textsubscript{2} during a 1~$\upmu$s \textit{npT} molecular dynamics simulation with the fitted Coulomb-Buckingham potential at 300~K and 1~bar (standard lab conditions). The top plot depicts the sodium coordination-environment by the angle $\theta$ over simulation time for each of the 12 simulated sodium layers as indicated by the structural models at the left and right side. $\theta$ is defined as the rotation angle of the upper coordinating oxygen ions to the lower ones as sketched in the Figure and explained for Figure~\ref{fig:Coordination}. The middle plot shows the mean-squared-displacement~(MSD) of sodium over simulation time along with the linear fits (dashed lines) for the regions of different intermediate phases. In the lower plot the sodium diffusion-coefficient resulting from the linear fits in the MSD plot is visualized over the simulation time.}
\end{figure*}

Finally, we performed a MD simulation at standard lab conditions (\textit{npT}, 300~K, and 1~bar) with our fitted Coulomb-Buckingham potential to study if it is possible to observe the O2 to P2 phase transition under experimental conditions. The results in Figure~\ref{fig:MD} indeed prove that the transition can be observed under standard lab conditions on the $\upmu$s-timescale. These timescales for a system with 12 layers (576 formula units, 2112 atoms) are only accessible by simple classical potentials highlighting the value of the Coulomb-Buckingham potential fitted in this work. Figure~\ref{fig:MD} shows that the O2 phase is kinetically stabilized for the first 100~ns. Next, one layer transitions and an O10P2 inter-growth phase remains stable for the next 250~ns. The rather long-time scale over which the thermodynamically unstable phases such as O2 and O10P2 exist under standard lab conditions indicates that kinetic phase stabilization can play an important role supporting the observation of "Z"(OP\textit{n})-phases by experiments as discussed before. Next, two more CoO\textsubscript{2} layers glide almost exactly at the same time resulting in a O6P6 phase with an OOPPOPPOOOPP stacking sequence. Finally, the transition of remaining O2 layers results in a P12 structure. However, before a complete transition, an O2P10 phase is observed for a short time (20~ns). The transition sequence is therefore O12-O10P2-O6P6-O2P10-P12. The sequence of phases observed during transition is symmetric and there is a tendency towards a decreased kinetic stability (shorter times to next transition) of the inter-growth phases towards a pure P-phase. Still, transition sequences are strongly temperature-dependent as a comparison with Figure~\ref{fig:Coordination}~(a) highlights and as discussed before most likely also random to some extend.

\subsection{Sodium Diffusion-Coefficients}

The $\upmu$s MD simulation trajectory also allows to analyse the sodium diffusivity directly at experimental conditions as many sodium hoppings occur during the simulation time. The mean squared displacement~(MSD) of sodium ions over simulation time is shown in Figure~\ref{fig:MD}. From the MSD plot, it is possible to determine the sodium diffusion coefficient ($D$) by~\cite{Usler.2023}:

\begin{equation}
    \label{eq:diffusion}
    D = \frac{1}{2d} \frac{\text{dMSD}}{\text{d}t}.
\end{equation}

Here, $d$ denotes the dimensionality of diffusion which is 2 in the case of 2D layered oxides and $\frac{\text{dMSD}}{\text{d}t}$ can be obtained as the slope of a linear fit in the MSD versus time plot with an \textit{y}-intercept of zero. The phase transitions significantly change the diffusivity of sodium ions as it is known that the P-phase has a better sodium conductivity (at lower sodium concentrations such as 0.67) than the O-phase counterpart due to a different diffusion mechanism~\cite{Mo.2014, Katcho.2017}. Thus, not a single linear fit of the MSD over the whole simulation time can be applied. Therefore, linear fits were performed for all different observed stacking sequences discussed before but omitting the O2P10 phase due to its short (in terms of simulation time) appearance. The linear fits are shown as dashed lines in the MSD plot in Figure~\ref{fig:MD}. Following Equation~\ref{eq:diffusion} the linear fits allow to determine the sodium diffusion coefficient and its evolution with the different stacking sequences is shown in Figure~\ref{fig:MD} as well.

The calculated sodium ion diffusion coefficients show a pronounced increase over simulation time as more and more P-layers form which is consistent with the fact that P-layers conduct sodium ions better than O-layers. For the pure P-phase (P12) finally a value of 3.92$\times$10$^{-7}$~cm\textsuperscript{2}/s is obtained which is in excellent agreement to an experimental study of single-crystal Na\textsubscript{\textit{x}}CoO\textsubscript{2} that reports a value of 1.2$\pm$0.5$\times$10$^{-7}$~cm\textsuperscript{2}/s at standard lab conditions and for sodium concentrations greater than 0.5~\cite{Shu.2008}. Other simulation based studies report values of 5.2$\times$10$^{-8}$~cm\textsuperscript{2}/s~\cite{Tatara.2025} (at a sodium concentration of 0.8) or approx. 5$\times$10$^{-6}$~cm\textsuperscript{2}/s~\cite{Mo.2014} (at 720~K) while further experimental studies on powders suggest values of 1.7$\times$10$^{-11}$~cm\textsuperscript{2}/s~\cite{Tatara.2025} or approx. 1$\times$10$^{-9}$~cm\textsuperscript{2}/s~\cite{Ohishi.2023}, and 1.9$\times$10$^{-11}$~cm\textsuperscript{2}/s~\cite{Shibata.2013}. It can be concluded that there is a broad variety sodium diffusion-coefficients reported in the literature for P2-Na\textsubscript{\textit{x}}CoO\textsubscript{2} while the value obtained in this work is very similar to the experimental value for single crystals. Thus, the Coulomb-Buckingham potential fitted in this work captures the sodium-ion dynamics reasonably well. Moreover, the simplicity of the potential allows to study the diffusion at standard lab conditions and over extremely long time scales ($\upmu$s). For the O2 phase experimental and theoretical data is harder to compare as the O2 phase is not commonly reported for Na\textsubscript{\textit{x}}CoO\textsubscript{2}. However, it remains interesting to compare the relative increase in the diffusion coefficient of the P2 structure versus the O2 structure. The fits in Figure~\ref{fig:MD} result in an diffusion coefficient of 1.90$\times$10$^{-8}$~cm\textsuperscript{2}/s for the pure O2 phase suggesting that the pure P2 phase shows an approx. 20-times increased sodium diffusion-coefficient. Recent simulation studies reported a 12-times higher sodium diffusion coefficient of P2-Na\textsubscript{0.8}CoO\textsubscript{2} over O3-Na\textsubscript{0.8}CoO\textsubscript{2}~\cite{Tatara.2025}. Moreover, at 720~K an (approximately) increase of 5-times in the P2 sodium diffusion coefficient over the O3 phase was reported~\cite{Mo.2014}. Another experimental study on Na\textsubscript{0.67}Fe\textsubscript{0.67}Mn\textsubscript{0.33}O\textsubscript{2} reports ratios of the sodium diffusion coefficient in the P2 over O3 structure in the range of 12-20 depending on the experimental method~\cite{Katcho.2017}. Overall, it can be concluded that the observed 20-times increase in the sodium diffusion-coefficient observed in the MD simulation in Figure~\ref{fig:MD} appears to be in a realistic order of magnitude and matches to values for similar systems reported in the literature supporting the conclusion that the Coulomb-Buckingham potential presented in this work is well-suited to study sodium-ion diffusion in Na\textsubscript{\textit{x}}CoO\textsubscript{2}, besides the phase transition.

\section{Conclusions}
In summary, we developed a Coulomb-Buckingham potential for layered oxide cathode materials at the example of Na\textsubscript{\textit{x}}CoO\textsubscript{2}. Our results show that most Buckingham parameters are insensitive to the charge-carrier (sodium) concentration ($x$) allowing for significant parameter reductions of the potential. We found that the most important parameter that changes with $x$ is the dielectric constant of the material to account for the change in screening of electrostatic interactions. However, the dependency of the dielectric constant with $x$ appears to be linear allowing for simple fittings and extrapolation. First, we computed the O2-P2 phase-transition barrier as function of $x$ by static calculations (single-point calculations) and moving one entire CoO\textsubscript{2} layer along the \textit{a}-lattice direction. A pronounced trend towards lower barriers at lower $x$ values and higher barriers for P2 to O2 than for O2 to P2 was found. The barrier values and trends were reasonably well reproduced by the fitted Coulomb-Buckingham potential. The Na-O Buckingham potential was identified to be the most important interaction beyond electrostatics to capture the phase-transition barriers. Afterwards, the O2 to P2 phase transition in Na\textsubscript{0.67}CoO\textsubscript{2} was studied by molecular dynamics simulations employing the fitted Coulomb-Buckingham potential. While it was shown that the transition mechanism during the simulations (shift of one entire CoO\textsubscript{2} layer) remains the same as in the static calculations, lower energy barriers were obtained by activation-energy calculations following Arrhenius equation. The fitted Coulomb-Buckingham potential has enabled performing molecular dynamics simulations at standard lab conditions on the timescale of $\upmu$s. The observed trends in gradual phase transitions (OP$n$-/"Z"-phases) and kinetic stabilities of intermediate phases might help in further understanding of experimental results on phase transitions and inter-growth phases. Finally, we computed the sodium-ion diffusion-coefficient as a function of completeness of phase transition. Absolute values of sodium diffusion-coefficients and the relative increase during phase transition were shown to be in similar order of magnitude than experimental values for single crystals reported in the literature.

Overall, this work shows, at the prototype layered oxide sodium-ion cathode material Na\textsubscript{\textit{x}}CoO\textsubscript{2}, that it is possible to investigate phase transitions and sodium dynamics at standard lab conditions utilizing simple pairwise Coulomb-Buckingham potentials. Further studies could consider different, more complex transition-metal compositions and include more layered phases such as O3 and P3. We believe that the procedure of fitting classical potentials outlined in this work can be a promising route to obtain more detailed understanding of unknown atomistic mechanisms in complex layered oxide sodium-ion cathode materials.

\begin{acknowledgments}
The authors gratefully acknowledge support by the “Deutsche Forschungsgemeinschaft” (DFG, German Research Foundation) in project No. 501562980. Moreover, computational resources were granted within Grant ID jiek12 by JARA-HPC on the JURECA~\cite{Thornig.2021} supercomputer at Forschugszentrum Jülich GmbH.
\end{acknowledgments}

\section*{Data Availability}
The raw data produced and analysed in this work is available from the corresponding author upon reasonable request.





\begin{thebibliography}{60}%
\makeatletter
\providecommand \@ifxundefined [1]{%
 \@ifx{#1\undefined}
}%
\providecommand \@ifnum [1]{%
 \ifnum #1\expandafter \@firstoftwo
 \else \expandafter \@secondoftwo
 \fi
}%
\providecommand \@ifx [1]{%
 \ifx #1\expandafter \@firstoftwo
 \else \expandafter \@secondoftwo
 \fi
}%
\providecommand \natexlab [1]{#1}%
\providecommand \enquote  [1]{``#1''}%
\providecommand \bibnamefont  [1]{#1}%
\providecommand \bibfnamefont [1]{#1}%
\providecommand \citenamefont [1]{#1}%
\providecommand \href@noop [0]{\@secondoftwo}%
\providecommand \href [0]{\begingroup \@sanitize@url \@href}%
\providecommand \@href[1]{\@@startlink{#1}\@@href}%
\providecommand \@@href[1]{\endgroup#1\@@endlink}%
\providecommand \@sanitize@url [0]{\catcode `\\12\catcode `\$12\catcode
  `\&12\catcode `\#12\catcode `\^12\catcode `\_12\catcode `\%12\relax}%
\providecommand \@@startlink[1]{}%
\providecommand \@@endlink[0]{}%
\providecommand \url  [0]{\begingroup\@sanitize@url \@url }%
\providecommand \@url [1]{\endgroup\@href {#1}{\urlprefix }}%
\providecommand \urlprefix  [0]{URL }%
\providecommand \Eprint [0]{\href }%
\providecommand \doibase [0]{https://doi.org/}%
\providecommand \selectlanguage [0]{\@gobble}%
\providecommand \bibinfo  [0]{\@secondoftwo}%
\providecommand \bibfield  [0]{\@secondoftwo}%
\providecommand \translation [1]{[#1]}%
\providecommand \BibitemOpen [0]{}%
\providecommand \bibitemStop [0]{}%
\providecommand \bibitemNoStop [0]{.\EOS\space}%
\providecommand \EOS [0]{\spacefactor3000\relax}%
\providecommand \BibitemShut  [1]{\csname bibitem#1\endcsname}%
\let\auto@bib@innerbib\@empty
\bibitem [{\citenamefont {Wang}\ \emph {et~al.}(2024)\citenamefont {Wang},
  \citenamefont {Zhu}, \citenamefont {Su}, \citenamefont {Guo}, \citenamefont
  {Chen}, \citenamefont {Liu}, \citenamefont {Dou}, \citenamefont {Chou},\ and\
  \citenamefont {Xiao}}]{Wang.2024}%
  \BibitemOpen
  \bibfield  {author} {\bibinfo {author} {\bibfnamefont {J.}~\bibnamefont
  {Wang}}, \bibinfo {author} {\bibfnamefont {Y.-F.}\ \bibnamefont {Zhu}},
  \bibinfo {author} {\bibfnamefont {Y.}~\bibnamefont {Su}}, \bibinfo {author}
  {\bibfnamefont {J.-X.}\ \bibnamefont {Guo}}, \bibinfo {author} {\bibfnamefont
  {S.}~\bibnamefont {Chen}}, \bibinfo {author} {\bibfnamefont {H.-K.}\
  \bibnamefont {Liu}}, \bibinfo {author} {\bibfnamefont {S.-X.}\ \bibnamefont
  {Dou}}, \bibinfo {author} {\bibfnamefont {S.-L.}\ \bibnamefont {Chou}},\ and\
  \bibinfo {author} {\bibfnamefont {Y.}~\bibnamefont {Xiao}},\ }\href
  {https://doi.org/10.1039/D3CS00929G} {\bibfield  {journal} {\bibinfo
  {journal} {Chemical Society reviews}\ }\textbf {\bibinfo {volume} {53}},\
  \bibinfo {pages} {4230} (\bibinfo {year} {2024})}\BibitemShut {NoStop}%
\bibitem [{\citenamefont {Yang}\ \emph {et~al.}(2024)\citenamefont {Yang},
  \citenamefont {Wang}, \citenamefont {Liu}, \citenamefont {Liu}, \citenamefont
  {Zhong}, \citenamefont {Song}, \citenamefont {Kong}, \citenamefont {Wu},\
  and\ \citenamefont {Guo}}]{Yang.2024}%
  \BibitemOpen
  \bibfield  {author} {\bibinfo {author} {\bibfnamefont {H.}~\bibnamefont
  {Yang}}, \bibinfo {author} {\bibfnamefont {D.}~\bibnamefont {Wang}}, \bibinfo
  {author} {\bibfnamefont {Y.}~\bibnamefont {Liu}}, \bibinfo {author}
  {\bibfnamefont {Y.}~\bibnamefont {Liu}}, \bibinfo {author} {\bibfnamefont
  {B.}~\bibnamefont {Zhong}}, \bibinfo {author} {\bibfnamefont
  {Y.}~\bibnamefont {Song}}, \bibinfo {author} {\bibfnamefont {Q.}~\bibnamefont
  {Kong}}, \bibinfo {author} {\bibfnamefont {Z.}~\bibnamefont {Wu}},\ and\
  \bibinfo {author} {\bibfnamefont {X.}~\bibnamefont {Guo}},\ }\href
  {https://doi.org/10.1039/D3EE02934D} {\bibfield  {journal} {\bibinfo
  {journal} {Energy {\&} Environmental Science}\ }\textbf {\bibinfo {volume}
  {17}},\ \bibinfo {pages} {1756} (\bibinfo {year} {2024})}\BibitemShut
  {NoStop}%
\bibitem [{\citenamefont {Wang}\ \emph {et~al.}(2018)\citenamefont {Wang},
  \citenamefont {You}, \citenamefont {Yin},\ and\ \citenamefont
  {Guo}}]{Wang.2018}%
  \BibitemOpen
  \bibfield  {author} {\bibinfo {author} {\bibfnamefont {P.-F.}\ \bibnamefont
  {Wang}}, \bibinfo {author} {\bibfnamefont {Y.}~\bibnamefont {You}}, \bibinfo
  {author} {\bibfnamefont {Y.-X.}\ \bibnamefont {Yin}},\ and\ \bibinfo {author}
  {\bibfnamefont {Y.-G.}\ \bibnamefont {Guo}},\ }\bibfield  {journal} {\bibinfo
   {journal} {Advanced Energy Materials}\ }\textbf {\bibinfo {volume} {8}},\
  \href {https://doi.org/10.1002/aenm.201701912} {10.1002/aenm.201701912}
  (\bibinfo {year} {2018})\BibitemShut {NoStop}%
\bibitem [{\citenamefont {Hwang}\ \emph {et~al.}(2017)\citenamefont {Hwang},
  \citenamefont {Myung},\ and\ \citenamefont {Sun}}]{Hwang.2017}%
  \BibitemOpen
  \bibfield  {author} {\bibinfo {author} {\bibfnamefont {J.-Y.}\ \bibnamefont
  {Hwang}}, \bibinfo {author} {\bibfnamefont {S.-T.}\ \bibnamefont {Myung}},\
  and\ \bibinfo {author} {\bibfnamefont {Y.-K.}\ \bibnamefont {Sun}},\ }\href
  {https://doi.org/10.1039/c6cs00776g} {\bibfield  {journal} {\bibinfo
  {journal} {Chemical Society reviews}\ }\textbf {\bibinfo {volume} {46}},\
  \bibinfo {pages} {3529} (\bibinfo {year} {2017})}\BibitemShut {NoStop}%
\bibitem [{\citenamefont {Zhang}\ and\ \citenamefont {Li}(2024)}]{Zhang.2024}%
  \BibitemOpen
  \bibfield  {author} {\bibinfo {author} {\bibfnamefont {J.}~\bibnamefont
  {Zhang}}\ and\ \bibinfo {author} {\bibfnamefont {X.}~\bibnamefont {Li}},\
  }\href {https://doi.org/10.1021/acs.energyfuels.4c02631} {\bibfield
  {journal} {\bibinfo  {journal} {Energy {\&} Fuels}\ }\textbf {\bibinfo
  {volume} {38}},\ \bibinfo {pages} {13906} (\bibinfo {year}
  {2024})}\BibitemShut {NoStop}%
\bibitem [{\citenamefont {Delmas}\ \emph {et~al.}(1980)\citenamefont {Delmas},
  \citenamefont {Fouassier},\ and\ \citenamefont {Hagenmuller}}]{Delmas.1980}%
  \BibitemOpen
  \bibfield  {author} {\bibinfo {author} {\bibfnamefont {C.}~\bibnamefont
  {Delmas}}, \bibinfo {author} {\bibfnamefont {C.}~\bibnamefont {Fouassier}},\
  and\ \bibinfo {author} {\bibfnamefont {P.}~\bibnamefont {Hagenmuller}},\
  }\href {https://doi.org/10.1016/0378-4363(80)90214-4} {\bibfield  {journal}
  {\bibinfo  {journal} {Physica B+C}\ }\textbf {\bibinfo {volume} {99}},\
  \bibinfo {pages} {81} (\bibinfo {year} {1980})}\BibitemShut {NoStop}%
\bibitem [{\citenamefont {Tang}\ \emph {et~al.}(2024)\citenamefont {Tang},
  \citenamefont {Zhang}, \citenamefont {Zuo}, \citenamefont {Zhou},
  \citenamefont {Zeng}, \citenamefont {Zhang}, \citenamefont {Zhang},
  \citenamefont {Huang}, \citenamefont {Zheng}, \citenamefont {Xu},
  \citenamefont {Yin}, \citenamefont {Qiu}, \citenamefont {Xiao}, \citenamefont
  {Zhang}, \citenamefont {Zhao}, \citenamefont {Liao}, \citenamefont {Hwang},
  \citenamefont {Sun}, \citenamefont {Amine}, \citenamefont {Wang},
  \citenamefont {Sun}, \citenamefont {Xu}, \citenamefont {Gu}, \citenamefont
  {Qiao},\ and\ \citenamefont {Sun}}]{Tang.2024}%
  \BibitemOpen
  \bibfield  {author} {\bibinfo {author} {\bibfnamefont {Y.}~\bibnamefont
  {Tang}}, \bibinfo {author} {\bibfnamefont {Q.}~\bibnamefont {Zhang}},
  \bibinfo {author} {\bibfnamefont {W.}~\bibnamefont {Zuo}}, \bibinfo {author}
  {\bibfnamefont {S.}~\bibnamefont {Zhou}}, \bibinfo {author} {\bibfnamefont
  {G.}~\bibnamefont {Zeng}}, \bibinfo {author} {\bibfnamefont {B.}~\bibnamefont
  {Zhang}}, \bibinfo {author} {\bibfnamefont {H.}~\bibnamefont {Zhang}},
  \bibinfo {author} {\bibfnamefont {Z.}~\bibnamefont {Huang}}, \bibinfo
  {author} {\bibfnamefont {L.}~\bibnamefont {Zheng}}, \bibinfo {author}
  {\bibfnamefont {J.}~\bibnamefont {Xu}}, \bibinfo {author} {\bibfnamefont
  {W.}~\bibnamefont {Yin}}, \bibinfo {author} {\bibfnamefont {Y.}~\bibnamefont
  {Qiu}}, \bibinfo {author} {\bibfnamefont {Y.}~\bibnamefont {Xiao}}, \bibinfo
  {author} {\bibfnamefont {Q.}~\bibnamefont {Zhang}}, \bibinfo {author}
  {\bibfnamefont {T.}~\bibnamefont {Zhao}}, \bibinfo {author} {\bibfnamefont
  {H.-G.}\ \bibnamefont {Liao}}, \bibinfo {author} {\bibfnamefont
  {I.}~\bibnamefont {Hwang}}, \bibinfo {author} {\bibfnamefont {C.-J.}\
  \bibnamefont {Sun}}, \bibinfo {author} {\bibfnamefont {K.}~\bibnamefont
  {Amine}}, \bibinfo {author} {\bibfnamefont {Q.}~\bibnamefont {Wang}},
  \bibinfo {author} {\bibfnamefont {Y.}~\bibnamefont {Sun}}, \bibinfo {author}
  {\bibfnamefont {G.-L.}\ \bibnamefont {Xu}}, \bibinfo {author} {\bibfnamefont
  {L.}~\bibnamefont {Gu}}, \bibinfo {author} {\bibfnamefont {Y.}~\bibnamefont
  {Qiao}},\ and\ \bibinfo {author} {\bibfnamefont {S.-G.}\ \bibnamefont
  {Sun}},\ }\href {https://doi.org/10.1038/s41893-024-01288-9} {\bibfield
  {journal} {\bibinfo  {journal} {Nature Sustainability}\ }\textbf {\bibinfo
  {volume} {7}},\ \bibinfo {pages} {348} (\bibinfo {year} {2024})}\BibitemShut
  {NoStop}%
\bibitem [{\citenamefont {Langella}\ \emph {et~al.}(2025)\citenamefont
  {Langella}, \citenamefont {Massaro}, \citenamefont {Mu{\~n}oz-Garc{\'i}a},\
  and\ \citenamefont {Pavone}}]{Langella.2025}%
  \BibitemOpen
  \bibfield  {author} {\bibinfo {author} {\bibfnamefont {A.}~\bibnamefont
  {Langella}}, \bibinfo {author} {\bibfnamefont {A.}~\bibnamefont {Massaro}},
  \bibinfo {author} {\bibfnamefont {A.~B.}\ \bibnamefont
  {Mu{\~n}oz-Garc{\'i}a}},\ and\ \bibinfo {author} {\bibfnamefont
  {M.}~\bibnamefont {Pavone}},\ }\href
  {https://doi.org/10.1021/acsenergylett.4c03335} {\bibfield  {journal}
  {\bibinfo  {journal} {ACS energy letters}\ }\textbf {\bibinfo {volume}
  {10}},\ \bibinfo {pages} {1089} (\bibinfo {year} {2025})}\BibitemShut
  {NoStop}%
\bibitem [{\citenamefont {Fellman}\ \emph {et~al.}(2025)\citenamefont
  {Fellman}, \citenamefont {Byggm{\"a}star}, \citenamefont {Granberg},
  \citenamefont {Nordlund},\ and\ \citenamefont {Djurabekova}}]{Fellman.2025}%
  \BibitemOpen
  \bibfield  {author} {\bibinfo {author} {\bibfnamefont {A.}~\bibnamefont
  {Fellman}}, \bibinfo {author} {\bibfnamefont {J.}~\bibnamefont
  {Byggm{\"a}star}}, \bibinfo {author} {\bibfnamefont {F.}~\bibnamefont
  {Granberg}}, \bibinfo {author} {\bibfnamefont {K.}~\bibnamefont {Nordlund}},\
  and\ \bibinfo {author} {\bibfnamefont {F.}~\bibnamefont {Djurabekova}},\
  }\bibfield  {journal} {\bibinfo  {journal} {Physical Review Materials}\
  }\textbf {\bibinfo {volume} {9}},\ \href
  {https://doi.org/10.1103/PhysRevMaterials.9.053807}
  {10.1103/PhysRevMaterials.9.053807} (\bibinfo {year} {2025})\BibitemShut
  {NoStop}%
\bibitem [{\citenamefont {Li}\ \emph {et~al.}(2025)\citenamefont {Li},
  \citenamefont {Zhang}, \citenamefont {Liu},\ and\ \citenamefont
  {Shen}}]{Li.2025}%
  \BibitemOpen
  \bibfield  {author} {\bibinfo {author} {\bibfnamefont {Y.}~\bibnamefont
  {Li}}, \bibinfo {author} {\bibfnamefont {X.}~\bibnamefont {Zhang}}, \bibinfo
  {author} {\bibfnamefont {M.}~\bibnamefont {Liu}},\ and\ \bibinfo {author}
  {\bibfnamefont {L.}~\bibnamefont {Shen}},\ }\bibfield  {journal} {\bibinfo
  {journal} {Journal of Materials Informatics}\ }\textbf {\bibinfo {volume}
  {5}},\ \href {https://doi.org/10.20517/jmi.2025.17} {10.20517/jmi.2025.17}
  (\bibinfo {year} {2025})\BibitemShut {NoStop}%
\bibitem [{\citenamefont {Deringer}\ \emph {et~al.}(2019)\citenamefont
  {Deringer}, \citenamefont {Caro},\ and\ \citenamefont
  {Cs{\'a}nyi}}]{Deringer.2019}%
  \BibitemOpen
  \bibfield  {author} {\bibinfo {author} {\bibfnamefont {V.~L.}\ \bibnamefont
  {Deringer}}, \bibinfo {author} {\bibfnamefont {M.~A.}\ \bibnamefont {Caro}},\
  and\ \bibinfo {author} {\bibfnamefont {G.}~\bibnamefont {Cs{\'a}nyi}},\
  }\href {https://doi.org/10.1002/adma.201902765} {\bibfield  {journal}
  {\bibinfo  {journal} {Advanced materials (Deerfield Beach, Fla.)}\ }\textbf
  {\bibinfo {volume} {31}},\ \bibinfo {pages} {e1902765} (\bibinfo {year}
  {2019})}\BibitemShut {NoStop}%
\bibitem [{\citenamefont {Behler}(2016)}]{Behler.2016}%
  \BibitemOpen
  \bibfield  {author} {\bibinfo {author} {\bibfnamefont {J.}~\bibnamefont
  {Behler}},\ }\href {https://doi.org/10.1063/1.4966192} {\bibfield  {journal}
  {\bibinfo  {journal} {The Journal of Chemical Physics}\ }\textbf {\bibinfo
  {volume} {145}},\ \bibinfo {pages} {170901} (\bibinfo {year}
  {2016})}\BibitemShut {NoStop}%
\bibitem [{\citenamefont {Mueller}\ \emph {et~al.}(2020)\citenamefont
  {Mueller}, \citenamefont {Hernandez},\ and\ \citenamefont
  {Wang}}]{Mueller.2020}%
  \BibitemOpen
  \bibfield  {author} {\bibinfo {author} {\bibfnamefont {T.}~\bibnamefont
  {Mueller}}, \bibinfo {author} {\bibfnamefont {A.}~\bibnamefont {Hernandez}},\
  and\ \bibinfo {author} {\bibfnamefont {C.}~\bibnamefont {Wang}},\ }\href
  {https://doi.org/10.1063/1.5126336} {\bibfield  {journal} {\bibinfo
  {journal} {The Journal of Chemical Physics}\ }\textbf {\bibinfo {volume}
  {152}},\ \bibinfo {pages} {050902} (\bibinfo {year} {2020})}\BibitemShut
  {NoStop}%
\bibitem [{\citenamefont {Lee}\ and\ \citenamefont {Park}(2012)}]{Lee.2012}%
  \BibitemOpen
  \bibfield  {author} {\bibinfo {author} {\bibfnamefont {S.}~\bibnamefont
  {Lee}}\ and\ \bibinfo {author} {\bibfnamefont {S.~S.}\ \bibnamefont {Park}},\
  }\href {https://doi.org/10.1021/jp2122467} {\bibfield  {journal} {\bibinfo
  {journal} {The Journal of Physical Chemistry C}\ }\textbf {\bibinfo {volume}
  {116}},\ \bibinfo {pages} {6484} (\bibinfo {year} {2012})}\BibitemShut
  {NoStop}%
\bibitem [{\citenamefont {Kerisit}\ \emph {et~al.}(2014)\citenamefont
  {Kerisit}, \citenamefont {Chaka}, \citenamefont {Droubay},\ and\
  \citenamefont {Ilton}}]{Kerisit.2014}%
  \BibitemOpen
  \bibfield  {author} {\bibinfo {author} {\bibfnamefont {S.}~\bibnamefont
  {Kerisit}}, \bibinfo {author} {\bibfnamefont {A.~M.}\ \bibnamefont {Chaka}},
  \bibinfo {author} {\bibfnamefont {T.~C.}\ \bibnamefont {Droubay}},\ and\
  \bibinfo {author} {\bibfnamefont {E.~S.}\ \bibnamefont {Ilton}},\ }\href
  {https://doi.org/10.1021/jp506025k} {\bibfield  {journal} {\bibinfo
  {journal} {The Journal of Physical Chemistry C}\ }\textbf {\bibinfo {volume}
  {118}},\ \bibinfo {pages} {24231} (\bibinfo {year} {2014})}\BibitemShut
  {NoStop}%
\bibitem [{\citenamefont {He}\ \emph {et~al.}(2019)\citenamefont {He},
  \citenamefont {Zhang},\ and\ \citenamefont {Liu}}]{He.2019}%
  \BibitemOpen
  \bibfield  {author} {\bibinfo {author} {\bibfnamefont {J.}~\bibnamefont
  {He}}, \bibinfo {author} {\bibfnamefont {L.}~\bibnamefont {Zhang}},\ and\
  \bibinfo {author} {\bibfnamefont {L.}~\bibnamefont {Liu}},\ }\href
  {https://doi.org/10.1039/c9cp01585j} {\bibfield  {journal} {\bibinfo
  {journal} {Physical chemistry chemical physics : PCCP}\ }\textbf {\bibinfo
  {volume} {21}},\ \bibinfo {pages} {12192} (\bibinfo {year}
  {2019})}\BibitemShut {NoStop}%
\bibitem [{\citenamefont {Hart}\ and\ \citenamefont {Bates}(1998)}]{Hart.1998}%
  \BibitemOpen
  \bibfield  {author} {\bibinfo {author} {\bibfnamefont {F.~X.}\ \bibnamefont
  {Hart}}\ and\ \bibinfo {author} {\bibfnamefont {J.~B.}\ \bibnamefont
  {Bates}},\ }\href {https://doi.org/10.1063/1.367521} {\bibfield  {journal}
  {\bibinfo  {journal} {Journal of Applied Physics}\ }\textbf {\bibinfo
  {volume} {83}},\ \bibinfo {pages} {7560} (\bibinfo {year}
  {1998})}\BibitemShut {NoStop}%
\bibitem [{\citenamefont {Morgan}\ \emph {et~al.}(2022)\citenamefont {Morgan},
  \citenamefont {Islam}, \citenamefont {Yang}, \citenamefont {O'Regan},
  \citenamefont {Patel}, \citenamefont {Ghosh}, \citenamefont {Kendrick},
  \citenamefont {Marinescu}, \citenamefont {Offer}, \citenamefont {Morgan},
  \citenamefont {Islam}, \citenamefont {Edge},\ and\ \citenamefont
  {Walsh}}]{Morgan.2022}%
  \BibitemOpen
  \bibfield  {author} {\bibinfo {author} {\bibfnamefont {L.~M.}\ \bibnamefont
  {Morgan}}, \bibinfo {author} {\bibfnamefont {M.~M.}\ \bibnamefont {Islam}},
  \bibinfo {author} {\bibfnamefont {H.}~\bibnamefont {Yang}}, \bibinfo {author}
  {\bibfnamefont {K.}~\bibnamefont {O'Regan}}, \bibinfo {author} {\bibfnamefont
  {A.~N.}\ \bibnamefont {Patel}}, \bibinfo {author} {\bibfnamefont
  {A.}~\bibnamefont {Ghosh}}, \bibinfo {author} {\bibfnamefont
  {E.}~\bibnamefont {Kendrick}}, \bibinfo {author} {\bibfnamefont
  {M.}~\bibnamefont {Marinescu}}, \bibinfo {author} {\bibfnamefont {G.~J.}\
  \bibnamefont {Offer}}, \bibinfo {author} {\bibfnamefont {B.~J.}\ \bibnamefont
  {Morgan}}, \bibinfo {author} {\bibfnamefont {M.~S.}\ \bibnamefont {Islam}},
  \bibinfo {author} {\bibfnamefont {J.}~\bibnamefont {Edge}},\ and\ \bibinfo
  {author} {\bibfnamefont {A.}~\bibnamefont {Walsh}},\ }\href
  {https://doi.org/10.1021/acsenergylett.1c02028} {\bibfield  {journal}
  {\bibinfo  {journal} {ACS energy letters}\ }\textbf {\bibinfo {volume} {7}},\
  \bibinfo {pages} {108} (\bibinfo {year} {2022})}\BibitemShut {NoStop}%
\bibitem [{\citenamefont {Sau}\ and\ \citenamefont
  {Ikeshoji}(2022{\natexlab{a}})}]{Sau.2022}%
  \BibitemOpen
  \bibfield  {author} {\bibinfo {author} {\bibfnamefont {K.}~\bibnamefont
  {Sau}}\ and\ \bibinfo {author} {\bibfnamefont {T.}~\bibnamefont {Ikeshoji}},\
  }\href {https://doi.org/10.1016/j.ssi.2022.115982} {\bibfield  {journal}
  {\bibinfo  {journal} {Solid State Ionics}\ }\textbf {\bibinfo {volume}
  {383}},\ \bibinfo {pages} {115982} (\bibinfo {year}
  {2022}{\natexlab{a}})}\BibitemShut {NoStop}%
\bibitem [{\citenamefont {Masese}\ \emph {et~al.}(2021)\citenamefont {Masese},
  \citenamefont {Miyazaki}, \citenamefont {Rizell}, \citenamefont {Kanyolo},
  \citenamefont {Chen}, \citenamefont {Ubukata}, \citenamefont {Kubota},
  \citenamefont {Sau}, \citenamefont {Ikeshoji}, \citenamefont {Huang},
  \citenamefont {Yoshii}, \citenamefont {Takahashi}, \citenamefont {Ito},
  \citenamefont {Senoh}, \citenamefont {Hwang}, \citenamefont {Alshehabi},
  \citenamefont {Matsumoto}, \citenamefont {Matsunaga}, \citenamefont {Fujii},
  \citenamefont {Yashima}, \citenamefont {Shikano}, \citenamefont {Tassel},
  \citenamefont {Kageyama}, \citenamefont {Uchimoto}, \citenamefont
  {Hagiwara},\ and\ \citenamefont {Saito}}]{Masese.2021}%
  \BibitemOpen
  \bibfield  {author} {\bibinfo {author} {\bibfnamefont {T.}~\bibnamefont
  {Masese}}, \bibinfo {author} {\bibfnamefont {Y.}~\bibnamefont {Miyazaki}},
  \bibinfo {author} {\bibfnamefont {J.}~\bibnamefont {Rizell}}, \bibinfo
  {author} {\bibfnamefont {G.~M.}\ \bibnamefont {Kanyolo}}, \bibinfo {author}
  {\bibfnamefont {C.-Y.}\ \bibnamefont {Chen}}, \bibinfo {author}
  {\bibfnamefont {H.}~\bibnamefont {Ubukata}}, \bibinfo {author} {\bibfnamefont
  {K.}~\bibnamefont {Kubota}}, \bibinfo {author} {\bibfnamefont
  {K.}~\bibnamefont {Sau}}, \bibinfo {author} {\bibfnamefont {T.}~\bibnamefont
  {Ikeshoji}}, \bibinfo {author} {\bibfnamefont {Z.-D.}\ \bibnamefont {Huang}},
  \bibinfo {author} {\bibfnamefont {K.}~\bibnamefont {Yoshii}}, \bibinfo
  {author} {\bibfnamefont {T.}~\bibnamefont {Takahashi}}, \bibinfo {author}
  {\bibfnamefont {M.}~\bibnamefont {Ito}}, \bibinfo {author} {\bibfnamefont
  {H.}~\bibnamefont {Senoh}}, \bibinfo {author} {\bibfnamefont
  {J.}~\bibnamefont {Hwang}}, \bibinfo {author} {\bibfnamefont
  {A.}~\bibnamefont {Alshehabi}}, \bibinfo {author} {\bibfnamefont
  {K.}~\bibnamefont {Matsumoto}}, \bibinfo {author} {\bibfnamefont
  {T.}~\bibnamefont {Matsunaga}}, \bibinfo {author} {\bibfnamefont
  {K.}~\bibnamefont {Fujii}}, \bibinfo {author} {\bibfnamefont
  {M.}~\bibnamefont {Yashima}}, \bibinfo {author} {\bibfnamefont
  {M.}~\bibnamefont {Shikano}}, \bibinfo {author} {\bibfnamefont
  {C.}~\bibnamefont {Tassel}}, \bibinfo {author} {\bibfnamefont
  {H.}~\bibnamefont {Kageyama}}, \bibinfo {author} {\bibfnamefont
  {Y.}~\bibnamefont {Uchimoto}}, \bibinfo {author} {\bibfnamefont
  {R.}~\bibnamefont {Hagiwara}},\ and\ \bibinfo {author} {\bibfnamefont
  {T.}~\bibnamefont {Saito}},\ }\href
  {https://doi.org/10.1038/s41467-021-24694-5} {\bibfield  {journal} {\bibinfo
  {journal} {Nature communications}\ }\textbf {\bibinfo {volume} {12}},\
  \bibinfo {pages} {4660} (\bibinfo {year} {2021})}\BibitemShut {NoStop}%
\bibitem [{\citenamefont {Sau}\ and\ \citenamefont
  {Ikeshoji}(2022{\natexlab{b}})}]{Sau.2022b}%
  \BibitemOpen
  \bibfield  {author} {\bibinfo {author} {\bibfnamefont {K.}~\bibnamefont
  {Sau}}\ and\ \bibinfo {author} {\bibfnamefont {T.}~\bibnamefont {Ikeshoji}},\
  }\bibfield  {journal} {\bibinfo  {journal} {Physical Review Materials}\
  }\textbf {\bibinfo {volume} {6}},\ \href
  {https://doi.org/10.1103/PhysRevMaterials.6.045406}
  {10.1103/PhysRevMaterials.6.045406} (\bibinfo {year}
  {2022}{\natexlab{b}})\BibitemShut {NoStop}%
\bibitem [{\citenamefont {Delmas}\ \emph {et~al.}(1981)\citenamefont {Delmas},
  \citenamefont {BRACONNIER}, \citenamefont {Fouassier},\ and\ \citenamefont
  {Hagenmuller}}]{Delmas.1981}%
  \BibitemOpen
  \bibfield  {author} {\bibinfo {author} {\bibfnamefont {C.}~\bibnamefont
  {Delmas}}, \bibinfo {author} {\bibfnamefont {J.}~\bibnamefont {BRACONNIER}},
  \bibinfo {author} {\bibfnamefont {C.}~\bibnamefont {Fouassier}},\ and\
  \bibinfo {author} {\bibfnamefont {P.}~\bibnamefont {Hagenmuller}},\ }\href
  {https://doi.org/10.1016/0167-2738(81)90076-X} {\bibfield  {journal}
  {\bibinfo  {journal} {Solid State Ionics}\ }\textbf {\bibinfo {volume}
  {3-4}},\ \bibinfo {pages} {165} (\bibinfo {year} {1981})}\BibitemShut
  {NoStop}%
\bibitem [{\citenamefont {Boddu}\ \emph {et~al.}(2021)\citenamefont {Boddu},
  \citenamefont {Puthusseri}, \citenamefont {Shirage}, \citenamefont {Mathur},\
  and\ \citenamefont {Pol}}]{Boddu.2021}%
  \BibitemOpen
  \bibfield  {author} {\bibinfo {author} {\bibfnamefont {V.~R.~R.}\
  \bibnamefont {Boddu}}, \bibinfo {author} {\bibfnamefont {D.}~\bibnamefont
  {Puthusseri}}, \bibinfo {author} {\bibfnamefont {P.~M.}\ \bibnamefont
  {Shirage}}, \bibinfo {author} {\bibfnamefont {P.}~\bibnamefont {Mathur}},\
  and\ \bibinfo {author} {\bibfnamefont {V.~G.}\ \bibnamefont {Pol}},\ }\href
  {https://doi.org/10.1007/s11581-021-04265-w} {\bibfield  {journal} {\bibinfo
  {journal} {Ionics}\ }\textbf {\bibinfo {volume} {27}},\ \bibinfo {pages}
  {4549} (\bibinfo {year} {2021})}\BibitemShut {NoStop}%
\bibitem [{\citenamefont {Kaufman}\ and\ \citenamefont {{van der
  Ven}}(2019)}]{Kaufman.2019}%
  \BibitemOpen
  \bibfield  {author} {\bibinfo {author} {\bibfnamefont {J.~L.}\ \bibnamefont
  {Kaufman}}\ and\ \bibinfo {author} {\bibfnamefont {A.}~\bibnamefont {{van der
  Ven}}},\ }\bibfield  {journal} {\bibinfo  {journal} {Physical Review
  Materials}\ }\textbf {\bibinfo {volume} {3}},\ \href
  {https://doi.org/10.1103/PhysRevMaterials.3.015402}
  {10.1103/PhysRevMaterials.3.015402} (\bibinfo {year} {2019})\BibitemShut
  {NoStop}%
\bibitem [{\citenamefont {Rivadulla}\ \emph {et~al.}(2003)\citenamefont
  {Rivadulla}, \citenamefont {Zhou},\ and\ \citenamefont
  {Goodenough}}]{Rivadulla.2003}%
  \BibitemOpen
  \bibfield  {author} {\bibinfo {author} {\bibfnamefont {F.}~\bibnamefont
  {Rivadulla}}, \bibinfo {author} {\bibfnamefont {J.-S.}\ \bibnamefont
  {Zhou}},\ and\ \bibinfo {author} {\bibfnamefont {J.~B.}\ \bibnamefont
  {Goodenough}},\ }\bibfield  {journal} {\bibinfo  {journal} {Physical review.
  B, Condensed matter}\ }\textbf {\bibinfo {volume} {68}},\ \href
  {https://doi.org/10.1103/PhysRevB.68.075108} {10.1103/PhysRevB.68.075108}
  (\bibinfo {year} {2003})\BibitemShut {NoStop}%
\bibitem [{\citenamefont {Toumar}\ \emph {et~al.}(2015)\citenamefont {Toumar},
  \citenamefont {Ong}, \citenamefont {Richards}, \citenamefont {Dacek},\ and\
  \citenamefont {Ceder}}]{Toumar.2015}%
  \BibitemOpen
  \bibfield  {author} {\bibinfo {author} {\bibfnamefont {A.~J.}\ \bibnamefont
  {Toumar}}, \bibinfo {author} {\bibfnamefont {S.~P.}\ \bibnamefont {Ong}},
  \bibinfo {author} {\bibfnamefont {W.~D.}\ \bibnamefont {Richards}}, \bibinfo
  {author} {\bibfnamefont {S.}~\bibnamefont {Dacek}},\ and\ \bibinfo {author}
  {\bibfnamefont {G.}~\bibnamefont {Ceder}},\ }\bibfield  {journal} {\bibinfo
  {journal} {Physical Review Applied}\ }\textbf {\bibinfo {volume} {4}},\ \href
  {https://doi.org/10.1103/PhysRevApplied.4.064002}
  {10.1103/PhysRevApplied.4.064002} (\bibinfo {year} {2015})\BibitemShut
  {NoStop}%
\bibitem [{\citenamefont {K{\"o}ster}\ and\ \citenamefont
  {Kaghazchi}(2024)}]{Koster.2024}%
  \BibitemOpen
  \bibfield  {author} {\bibinfo {author} {\bibfnamefont {K.}~\bibnamefont
  {K{\"o}ster}}\ and\ \bibinfo {author} {\bibfnamefont {P.}~\bibnamefont
  {Kaghazchi}},\ }\bibfield  {journal} {\bibinfo  {journal} {Physical Review
  B}\ }\textbf {\bibinfo {volume} {109}},\ \href
  {https://doi.org/10.1103/PhysRevB.109.155134} {10.1103/PhysRevB.109.155134}
  (\bibinfo {year} {2024})\BibitemShut {NoStop}%
\bibitem [{\citenamefont {Biecher}\ \emph {et~al.}(2022)\citenamefont
  {Biecher}, \citenamefont {Baux}, \citenamefont {Fauth}, \citenamefont
  {Delmas}, \citenamefont {Goward},\ and\ \citenamefont
  {Carlier}}]{Biecher.2022}%
  \BibitemOpen
  \bibfield  {author} {\bibinfo {author} {\bibfnamefont {Y.}~\bibnamefont
  {Biecher}}, \bibinfo {author} {\bibfnamefont {A.}~\bibnamefont {Baux}},
  \bibinfo {author} {\bibfnamefont {F.}~\bibnamefont {Fauth}}, \bibinfo
  {author} {\bibfnamefont {C.}~\bibnamefont {Delmas}}, \bibinfo {author}
  {\bibfnamefont {G.~R.}\ \bibnamefont {Goward}},\ and\ \bibinfo {author}
  {\bibfnamefont {D.}~\bibnamefont {Carlier}},\ }\href
  {https://doi.org/10.1021/acs.chemmater.2c01055} {\bibfield  {journal}
  {\bibinfo  {journal} {Chemistry of Materials}\ }\textbf {\bibinfo {volume}
  {34}},\ \bibinfo {pages} {6431} (\bibinfo {year} {2022})}\BibitemShut
  {NoStop}%
\bibitem [{\citenamefont {Lee}\ \emph {et~al.}(2013)\citenamefont {Lee},
  \citenamefont {Xu},\ and\ \citenamefont {Meng}}]{Lee.2013}%
  \BibitemOpen
  \bibfield  {author} {\bibinfo {author} {\bibfnamefont {D.~H.}\ \bibnamefont
  {Lee}}, \bibinfo {author} {\bibfnamefont {J.}~\bibnamefont {Xu}},\ and\
  \bibinfo {author} {\bibfnamefont {Y.~S.}\ \bibnamefont {Meng}},\ }\href
  {https://doi.org/10.1039/c2cp44467d} {\bibfield  {journal} {\bibinfo
  {journal} {Physical chemistry chemical physics : PCCP}\ }\textbf {\bibinfo
  {volume} {15}},\ \bibinfo {pages} {3304} (\bibinfo {year}
  {2013})}\BibitemShut {NoStop}%
\bibitem [{\citenamefont {Ewald}(1921)}]{Ewald.1921}%
  \BibitemOpen
  \bibfield  {author} {\bibinfo {author} {\bibfnamefont {P.~P.}\ \bibnamefont
  {Ewald}},\ }\href {https://doi.org/10.1002/andp.19213690304} {\bibfield
  {journal} {\bibinfo  {journal} {Annalen der Physik}\ }\textbf {\bibinfo
  {volume} {369}},\ \bibinfo {pages} {253} (\bibinfo {year}
  {1921})}\BibitemShut {NoStop}%
\bibitem [{\citenamefont {Buckingham}(1938)}]{Buckingham.1938}%
  \BibitemOpen
  \bibfield  {author} {\bibinfo {author} {\bibfnamefont {R.~A.}\ \bibnamefont
  {Buckingham}},\ }\href {https://doi.org/10.1098/rspa.1938.0173} {\bibfield
  {journal} {\bibinfo  {journal} {Proceedings of the Royal Society of London.
  Series A. Mathematical and Physical Sciences}\ }\textbf {\bibinfo {volume}
  {168}},\ \bibinfo {pages} {264} (\bibinfo {year} {1938})}\BibitemShut
  {NoStop}%
\bibitem [{\citenamefont {K{\"o}ster}\ \emph {et~al.}(2025)\citenamefont
  {K{\"o}ster}, \citenamefont {Binninger},\ and\ \citenamefont
  {Kaghazchi}}]{Koster.2025}%
  \BibitemOpen
  \bibfield  {author} {\bibinfo {author} {\bibfnamefont {K.}~\bibnamefont
  {K{\"o}ster}}, \bibinfo {author} {\bibfnamefont {T.}~\bibnamefont
  {Binninger}},\ and\ \bibinfo {author} {\bibfnamefont {P.}~\bibnamefont
  {Kaghazchi}},\ }\bibfield  {journal} {\bibinfo  {journal} {npj Computational
  Materials}\ }\textbf {\bibinfo {volume} {11}},\ \href
  {https://doi.org/10.1038/s41524-025-01690-7} {10.1038/s41524-025-01690-7}
  (\bibinfo {year} {2025})\BibitemShut {NoStop}%
\bibitem [{\citenamefont {Bl{\"o}chl}(1994)}]{Blochl.1994}%
  \BibitemOpen
  \bibfield  {author} {\bibinfo {author} {\bibfnamefont {P.~E.}\ \bibnamefont
  {Bl{\"o}chl}},\ }\href {https://doi.org/10.1103/physrevb.50.17953} {\bibfield
   {journal} {\bibinfo  {journal} {Physical review. B, Condensed matter}\
  }\textbf {\bibinfo {volume} {50}},\ \bibinfo {pages} {17953} (\bibinfo {year}
  {1994})}\BibitemShut {NoStop}%
\bibitem [{\citenamefont {Kresse}\ and\ \citenamefont
  {Furthm{\"u}ller}(1996)}]{Kresse.1996}%
  \BibitemOpen
  \bibfield  {author} {\bibinfo {author} {\bibfnamefont {G.}~\bibnamefont
  {Kresse}}\ and\ \bibinfo {author} {\bibfnamefont {J.}~\bibnamefont
  {Furthm{\"u}ller}},\ }\href {https://doi.org/10.1103/physrevb.54.11169}
  {\bibfield  {journal} {\bibinfo  {journal} {Physical review. B, Condensed
  matter}\ }\textbf {\bibinfo {volume} {54}},\ \bibinfo {pages} {11169}
  (\bibinfo {year} {1996})}\BibitemShut {NoStop}%
\bibitem [{\citenamefont {Perdew}\ \emph {et~al.}(1996)\citenamefont {Perdew},
  \citenamefont {Burke},\ and\ \citenamefont {Ernzerhof}}]{Perdew.1996}%
  \BibitemOpen
  \bibfield  {author} {\bibinfo {author} {\bibfnamefont {J.~P.}\ \bibnamefont
  {Perdew}}, \bibinfo {author} {\bibfnamefont {K.}~\bibnamefont {Burke}},\ and\
  \bibinfo {author} {\bibfnamefont {M.}~\bibnamefont {Ernzerhof}},\ }\href
  {https://doi.org/10.1103/PhysRevLett.77.3865} {\bibfield  {journal} {\bibinfo
   {journal} {Physical review letters}\ }\textbf {\bibinfo {volume} {77}},\
  \bibinfo {pages} {3865} (\bibinfo {year} {1996})}\BibitemShut {NoStop}%
\bibitem [{\citenamefont {Grimme}(2006)}]{Grimme.2006}%
  \BibitemOpen
  \bibfield  {author} {\bibinfo {author} {\bibfnamefont {S.}~\bibnamefont
  {Grimme}},\ }\href {https://doi.org/10.1002/jcc.20495} {\bibfield  {journal}
  {\bibinfo  {journal} {Journal of computational chemistry}\ }\textbf {\bibinfo
  {volume} {27}},\ \bibinfo {pages} {1787} (\bibinfo {year}
  {2006})}\BibitemShut {NoStop}%
\bibitem [{\citenamefont {Virtanen}\ \emph {et~al.}(2020)\citenamefont
  {Virtanen}, \citenamefont {Gommers}, \citenamefont {Oliphant}, \citenamefont
  {Haberland}, \citenamefont {Reddy}, \citenamefont {Cournapeau}, \citenamefont
  {Burovski}, \citenamefont {Peterson}, \citenamefont {Weckesser},
  \citenamefont {Bright}, \citenamefont {{van der Walt}}, \citenamefont
  {Brett}, \citenamefont {Wilson}, \citenamefont {Millman}, \citenamefont
  {Mayorov}, \citenamefont {Nelson}, \citenamefont {Jones}, \citenamefont
  {Kern}, \citenamefont {Larson}, \citenamefont {Carey}, \citenamefont {Polat},
  \citenamefont {Feng}, \citenamefont {Moore}, \citenamefont {VanderPlas},
  \citenamefont {Laxalde}, \citenamefont {Perktold}, \citenamefont {Cimrman},
  \citenamefont {Henriksen}, \citenamefont {Quintero}, \citenamefont {Harris},
  \citenamefont {Archibald}, \citenamefont {Ribeiro}, \citenamefont
  {Pedregosa},\ and\ \citenamefont {{van Mulbregt}}}]{Virtanen.2020}%
  \BibitemOpen
  \bibfield  {author} {\bibinfo {author} {\bibfnamefont {P.}~\bibnamefont
  {Virtanen}}, \bibinfo {author} {\bibfnamefont {R.}~\bibnamefont {Gommers}},
  \bibinfo {author} {\bibfnamefont {T.~E.}\ \bibnamefont {Oliphant}}, \bibinfo
  {author} {\bibfnamefont {M.}~\bibnamefont {Haberland}}, \bibinfo {author}
  {\bibfnamefont {T.}~\bibnamefont {Reddy}}, \bibinfo {author} {\bibfnamefont
  {D.}~\bibnamefont {Cournapeau}}, \bibinfo {author} {\bibfnamefont
  {E.}~\bibnamefont {Burovski}}, \bibinfo {author} {\bibfnamefont
  {P.}~\bibnamefont {Peterson}}, \bibinfo {author} {\bibfnamefont
  {W.}~\bibnamefont {Weckesser}}, \bibinfo {author} {\bibfnamefont
  {J.}~\bibnamefont {Bright}}, \bibinfo {author} {\bibfnamefont {S.~J.}\
  \bibnamefont {{van der Walt}}}, \bibinfo {author} {\bibfnamefont
  {M.}~\bibnamefont {Brett}}, \bibinfo {author} {\bibfnamefont
  {J.}~\bibnamefont {Wilson}}, \bibinfo {author} {\bibfnamefont {K.~J.}\
  \bibnamefont {Millman}}, \bibinfo {author} {\bibfnamefont {N.}~\bibnamefont
  {Mayorov}}, \bibinfo {author} {\bibfnamefont {A.~R.~J.}\ \bibnamefont
  {Nelson}}, \bibinfo {author} {\bibfnamefont {E.}~\bibnamefont {Jones}},
  \bibinfo {author} {\bibfnamefont {R.}~\bibnamefont {Kern}}, \bibinfo {author}
  {\bibfnamefont {E.}~\bibnamefont {Larson}}, \bibinfo {author} {\bibfnamefont
  {C.~J.}\ \bibnamefont {Carey}}, \bibinfo {author} {\bibfnamefont
  {{\.{I}}.}~\bibnamefont {Polat}}, \bibinfo {author} {\bibfnamefont
  {Y.}~\bibnamefont {Feng}}, \bibinfo {author} {\bibfnamefont {E.~W.}\
  \bibnamefont {Moore}}, \bibinfo {author} {\bibfnamefont {J.}~\bibnamefont
  {VanderPlas}}, \bibinfo {author} {\bibfnamefont {D.}~\bibnamefont {Laxalde}},
  \bibinfo {author} {\bibfnamefont {J.}~\bibnamefont {Perktold}}, \bibinfo
  {author} {\bibfnamefont {R.}~\bibnamefont {Cimrman}}, \bibinfo {author}
  {\bibfnamefont {I.}~\bibnamefont {Henriksen}}, \bibinfo {author}
  {\bibfnamefont {E.~A.}\ \bibnamefont {Quintero}}, \bibinfo {author}
  {\bibfnamefont {C.~R.}\ \bibnamefont {Harris}}, \bibinfo {author}
  {\bibfnamefont {A.~M.}\ \bibnamefont {Archibald}}, \bibinfo {author}
  {\bibfnamefont {A.~H.}\ \bibnamefont {Ribeiro}}, \bibinfo {author}
  {\bibfnamefont {F.}~\bibnamefont {Pedregosa}},\ and\ \bibinfo {author}
  {\bibfnamefont {P.}~\bibnamefont {{van Mulbregt}}},\ }\href
  {https://doi.org/10.1038/s41592-019-0686-2} {\bibfield  {journal} {\bibinfo
  {journal} {Nature methods}\ }\textbf {\bibinfo {volume} {17}},\ \bibinfo
  {pages} {261} (\bibinfo {year} {2020})}\BibitemShut {NoStop}%
\bibitem [{\citenamefont {Storn}\ and\ \citenamefont
  {Price}(1997)}]{Storn.1997}%
  \BibitemOpen
  \bibfield  {author} {\bibinfo {author} {\bibfnamefont {R.}~\bibnamefont
  {Storn}}\ and\ \bibinfo {author} {\bibfnamefont {K.}~\bibnamefont {Price}},\
  }\href {https://doi.org/10.1023/A:1008202821328} {\bibfield  {journal}
  {\bibinfo  {journal} {Journal of Global Optimization}\ }\textbf {\bibinfo
  {volume} {11}},\ \bibinfo {pages} {341} (\bibinfo {year} {1997})}\BibitemShut
  {NoStop}%
\bibitem [{\citenamefont {Liu}\ and\ \citenamefont {Nocedal}(1989)}]{Liu.1989}%
  \BibitemOpen
  \bibfield  {author} {\bibinfo {author} {\bibfnamefont {D.~C.}\ \bibnamefont
  {Liu}}\ and\ \bibinfo {author} {\bibfnamefont {J.}~\bibnamefont {Nocedal}},\
  }\href {https://doi.org/10.1007/BF01589116} {\bibfield  {journal} {\bibinfo
  {journal} {Mathematical Programming}\ }\textbf {\bibinfo {volume} {45}},\
  \bibinfo {pages} {503} (\bibinfo {year} {1989})}\BibitemShut {NoStop}%
\bibitem [{\citenamefont {Li}\ \emph {et~al.}(2004)\citenamefont {Li},
  \citenamefont {Yang}, \citenamefont {Hou},\ and\ \citenamefont
  {Zhu}}]{Li.2004}%
  \BibitemOpen
  \bibfield  {author} {\bibinfo {author} {\bibfnamefont {Z.}~\bibnamefont
  {Li}}, \bibinfo {author} {\bibfnamefont {J.}~\bibnamefont {Yang}}, \bibinfo
  {author} {\bibfnamefont {J.~G.}\ \bibnamefont {Hou}},\ and\ \bibinfo {author}
  {\bibfnamefont {Q.}~\bibnamefont {Zhu}},\ }\bibfield  {journal} {\bibinfo
  {journal} {Physical Review B}\ }\textbf {\bibinfo {volume} {70}},\ \href
  {https://doi.org/10.1103/PhysRevB.70.144518} {10.1103/PhysRevB.70.144518}
  (\bibinfo {year} {2004})\BibitemShut {NoStop}%
\bibitem [{\citenamefont {Petousis}\ \emph {et~al.}(2017)\citenamefont
  {Petousis}, \citenamefont {Mrdjenovich}, \citenamefont {Ballouz},
  \citenamefont {Liu}, \citenamefont {Winston}, \citenamefont {Chen},
  \citenamefont {Graf}, \citenamefont {Schladt}, \citenamefont {Persson},\ and\
  \citenamefont {Prinz}}]{Petousis.2017}%
  \BibitemOpen
  \bibfield  {author} {\bibinfo {author} {\bibfnamefont {I.}~\bibnamefont
  {Petousis}}, \bibinfo {author} {\bibfnamefont {D.}~\bibnamefont
  {Mrdjenovich}}, \bibinfo {author} {\bibfnamefont {E.}~\bibnamefont
  {Ballouz}}, \bibinfo {author} {\bibfnamefont {M.}~\bibnamefont {Liu}},
  \bibinfo {author} {\bibfnamefont {D.}~\bibnamefont {Winston}}, \bibinfo
  {author} {\bibfnamefont {W.}~\bibnamefont {Chen}}, \bibinfo {author}
  {\bibfnamefont {T.}~\bibnamefont {Graf}}, \bibinfo {author} {\bibfnamefont
  {T.~D.}\ \bibnamefont {Schladt}}, \bibinfo {author} {\bibfnamefont {K.~A.}\
  \bibnamefont {Persson}},\ and\ \bibinfo {author} {\bibfnamefont {F.~B.}\
  \bibnamefont {Prinz}},\ }\href {https://doi.org/10.1038/sdata.2016.134}
  {\bibfield  {journal} {\bibinfo  {journal} {Scientific data}\ }\textbf
  {\bibinfo {volume} {4}},\ \bibinfo {pages} {160134} (\bibinfo {year}
  {2017})}\BibitemShut {NoStop}%
\bibitem [{\citenamefont {Thompson}\ \emph {et~al.}(2022)\citenamefont
  {Thompson}, \citenamefont {Aktulga}, \citenamefont {Berger}, \citenamefont
  {Bolintineanu}, \citenamefont {Brown}, \citenamefont {Crozier}, \citenamefont
  {in~'t Veld}, \citenamefont {Kohlmeyer}, \citenamefont {Moore}, \citenamefont
  {Nguyen}, \citenamefont {Shan}, \citenamefont {Stevens}, \citenamefont
  {Tranchida}, \citenamefont {Trott},\ and\ \citenamefont
  {Plimpton}}]{Thompson.2022}%
  \BibitemOpen
  \bibfield  {author} {\bibinfo {author} {\bibfnamefont {A.~P.}\ \bibnamefont
  {Thompson}}, \bibinfo {author} {\bibfnamefont {H.~M.}\ \bibnamefont
  {Aktulga}}, \bibinfo {author} {\bibfnamefont {R.}~\bibnamefont {Berger}},
  \bibinfo {author} {\bibfnamefont {D.~S.}\ \bibnamefont {Bolintineanu}},
  \bibinfo {author} {\bibfnamefont {W.~M.}\ \bibnamefont {Brown}}, \bibinfo
  {author} {\bibfnamefont {P.~S.}\ \bibnamefont {Crozier}}, \bibinfo {author}
  {\bibfnamefont {P.~J.}\ \bibnamefont {in~'t Veld}}, \bibinfo {author}
  {\bibfnamefont {A.}~\bibnamefont {Kohlmeyer}}, \bibinfo {author}
  {\bibfnamefont {S.~G.}\ \bibnamefont {Moore}}, \bibinfo {author}
  {\bibfnamefont {T.~D.}\ \bibnamefont {Nguyen}}, \bibinfo {author}
  {\bibfnamefont {R.}~\bibnamefont {Shan}}, \bibinfo {author} {\bibfnamefont
  {M.~J.}\ \bibnamefont {Stevens}}, \bibinfo {author} {\bibfnamefont
  {J.}~\bibnamefont {Tranchida}}, \bibinfo {author} {\bibfnamefont
  {C.}~\bibnamefont {Trott}},\ and\ \bibinfo {author} {\bibfnamefont {S.~J.}\
  \bibnamefont {Plimpton}},\ }\href {https://doi.org/10.1016/j.cpc.2021.108171}
  {\bibfield  {journal} {\bibinfo  {journal} {Computer Physics Communications}\
  }\textbf {\bibinfo {volume} {271}},\ \bibinfo {pages} {108171} (\bibinfo
  {year} {2022})}\BibitemShut {NoStop}%
\bibitem [{\citenamefont {Plimpton}(1995)}]{Plimpton.1995}%
  \BibitemOpen
  \bibfield  {author} {\bibinfo {author} {\bibfnamefont {S.}~\bibnamefont
  {Plimpton}},\ }\href {https://doi.org/10.1006/jcph.1995.1039} {\bibfield
  {journal} {\bibinfo  {journal} {Journal of Computational Physics}\ }\textbf
  {\bibinfo {volume} {117}},\ \bibinfo {pages} {1} (\bibinfo {year}
  {1995})}\BibitemShut {NoStop}%
\bibitem [{\citenamefont {Bitzek}\ \emph {et~al.}(2006)\citenamefont {Bitzek},
  \citenamefont {Koskinen}, \citenamefont {G{\"a}hler}, \citenamefont
  {Moseler},\ and\ \citenamefont {Gumbsch}}]{Bitzek.2006}%
  \BibitemOpen
  \bibfield  {author} {\bibinfo {author} {\bibfnamefont {E.}~\bibnamefont
  {Bitzek}}, \bibinfo {author} {\bibfnamefont {P.}~\bibnamefont {Koskinen}},
  \bibinfo {author} {\bibfnamefont {F.}~\bibnamefont {G{\"a}hler}}, \bibinfo
  {author} {\bibfnamefont {M.}~\bibnamefont {Moseler}},\ and\ \bibinfo {author}
  {\bibfnamefont {P.}~\bibnamefont {Gumbsch}},\ }\href
  {https://doi.org/10.1103/PhysRevLett.97.170201} {\bibfield  {journal}
  {\bibinfo  {journal} {Physical review letters}\ }\textbf {\bibinfo {volume}
  {97}},\ \bibinfo {pages} {170201} (\bibinfo {year} {2006})}\BibitemShut
  {NoStop}%
\bibitem [{\citenamefont {Nos{\'e}}(1984)}]{Nose.1984}%
  \BibitemOpen
  \bibfield  {author} {\bibinfo {author} {\bibfnamefont {S.}~\bibnamefont
  {Nos{\'e}}},\ }\href {https://doi.org/10.1063/1.447334} {\bibfield  {journal}
  {\bibinfo  {journal} {The Journal of Chemical Physics}\ }\textbf {\bibinfo
  {volume} {81}},\ \bibinfo {pages} {511} (\bibinfo {year} {1984})}\BibitemShut
  {NoStop}%
\bibitem [{\citenamefont {Hoover}(1985)}]{Hoover.1985}%
  \BibitemOpen
  \bibfield  {author} {\bibinfo {author} {\bibfnamefont {W.~G.}\ \bibnamefont
  {Hoover}},\ }\href {https://doi.org/10.1103/physreva.31.1695} {\bibfield
  {journal} {\bibinfo  {journal} {Physical review. A, General physics}\
  }\textbf {\bibinfo {volume} {31}},\ \bibinfo {pages} {1695} (\bibinfo {year}
  {1985})}\BibitemShut {NoStop}%
\bibitem [{\citenamefont {Shinoda}\ \emph {et~al.}(2004)\citenamefont
  {Shinoda}, \citenamefont {Shiga},\ and\ \citenamefont
  {Mikami}}]{Shinoda.2004}%
  \BibitemOpen
  \bibfield  {author} {\bibinfo {author} {\bibfnamefont {W.}~\bibnamefont
  {Shinoda}}, \bibinfo {author} {\bibfnamefont {M.}~\bibnamefont {Shiga}},\
  and\ \bibinfo {author} {\bibfnamefont {M.}~\bibnamefont {Mikami}},\
  }\bibfield  {journal} {\bibinfo  {journal} {Physical Review B}\ }\textbf
  {\bibinfo {volume} {69}},\ \href {https://doi.org/10.1103/PhysRevB.69.134103}
  {10.1103/PhysRevB.69.134103} (\bibinfo {year} {2004})\BibitemShut {NoStop}%
\bibitem [{\citenamefont {Tuckerman}\ \emph {et~al.}(2006)\citenamefont
  {Tuckerman}, \citenamefont {Alejandre}, \citenamefont {L{\'o}pez-Rend{\'o}n},
  \citenamefont {Jochim},\ and\ \citenamefont {Martyna}}]{Tuckerman.2006}%
  \BibitemOpen
  \bibfield  {author} {\bibinfo {author} {\bibfnamefont {M.~E.}\ \bibnamefont
  {Tuckerman}}, \bibinfo {author} {\bibfnamefont {J.}~\bibnamefont
  {Alejandre}}, \bibinfo {author} {\bibfnamefont {R.}~\bibnamefont
  {L{\'o}pez-Rend{\'o}n}}, \bibinfo {author} {\bibfnamefont {A.~L.}\
  \bibnamefont {Jochim}},\ and\ \bibinfo {author} {\bibfnamefont {G.~J.}\
  \bibnamefont {Martyna}},\ }\href
  {https://doi.org/10.1088/0305-4470/39/19/S18} {\bibfield  {journal} {\bibinfo
   {journal} {Journal of Physics A: Mathematical and General}\ }\textbf
  {\bibinfo {volume} {39}},\ \bibinfo {pages} {5629} (\bibinfo {year}
  {2006})}\BibitemShut {NoStop}%
\bibitem [{\citenamefont {Momma}\ and\ \citenamefont
  {Izumi}(2008)}]{Momma.2008}%
  \BibitemOpen
  \bibfield  {author} {\bibinfo {author} {\bibfnamefont {K.}~\bibnamefont
  {Momma}}\ and\ \bibinfo {author} {\bibfnamefont {F.}~\bibnamefont {Izumi}},\
  }\href {https://doi.org/10.1107/S0021889808012016} {\bibfield  {journal}
  {\bibinfo  {journal} {Journal of Applied Crystallography}\ }\textbf {\bibinfo
  {volume} {41}},\ \bibinfo {pages} {653} (\bibinfo {year} {2008})}\BibitemShut
  {NoStop}%
\bibitem [{\citenamefont {Berthelot}\ \emph {et~al.}(2011)\citenamefont
  {Berthelot}, \citenamefont {Carlier},\ and\ \citenamefont
  {Delmas}}]{Berthelot.2011}%
  \BibitemOpen
  \bibfield  {author} {\bibinfo {author} {\bibfnamefont {R.}~\bibnamefont
  {Berthelot}}, \bibinfo {author} {\bibfnamefont {D.}~\bibnamefont {Carlier}},\
  and\ \bibinfo {author} {\bibfnamefont {C.}~\bibnamefont {Delmas}},\ }\href
  {https://doi.org/10.1038/nmat2920} {\bibfield  {journal} {\bibinfo  {journal}
  {Nature materials}\ }\textbf {\bibinfo {volume} {10}},\ \bibinfo {pages} {74}
  (\bibinfo {year} {2011})}\BibitemShut {NoStop}%
\bibitem [{\citenamefont {{Mortemard de Boisse}}\ \emph
  {et~al.}(2014)\citenamefont {{Mortemard de Boisse}}, \citenamefont {Carlier},
  \citenamefont {Guignard}, \citenamefont {Bourgeois},\ and\ \citenamefont
  {Delmas}}]{MortemarddeBoisse.2014}%
  \BibitemOpen
  \bibfield  {author} {\bibinfo {author} {\bibfnamefont {B.}~\bibnamefont
  {{Mortemard de Boisse}}}, \bibinfo {author} {\bibfnamefont {D.}~\bibnamefont
  {Carlier}}, \bibinfo {author} {\bibfnamefont {M.}~\bibnamefont {Guignard}},
  \bibinfo {author} {\bibfnamefont {L.}~\bibnamefont {Bourgeois}},\ and\
  \bibinfo {author} {\bibfnamefont {C.}~\bibnamefont {Delmas}},\ }\href
  {https://doi.org/10.1021/ic5017802} {\bibfield  {journal} {\bibinfo
  {journal} {Inorganic chemistry}\ }\textbf {\bibinfo {volume} {53}},\ \bibinfo
  {pages} {11197} (\bibinfo {year} {2014})}\BibitemShut {NoStop}%
\bibitem [{\citenamefont {Somerville}\ \emph {et~al.}(2019)\citenamefont
  {Somerville}, \citenamefont {Sobkowiak}, \citenamefont {Tapia-Ruiz},
  \citenamefont {Billaud}, \citenamefont {Lozano}, \citenamefont {House},
  \citenamefont {Gallington}, \citenamefont {Ericsson}, \citenamefont
  {H{\"a}ggstr{\"o}m}, \citenamefont {Roberts}, \citenamefont {Maitra},\ and\
  \citenamefont {Bruce}}]{Somerville.2019}%
  \BibitemOpen
  \bibfield  {author} {\bibinfo {author} {\bibfnamefont {J.~W.}\ \bibnamefont
  {Somerville}}, \bibinfo {author} {\bibfnamefont {A.}~\bibnamefont
  {Sobkowiak}}, \bibinfo {author} {\bibfnamefont {N.}~\bibnamefont
  {Tapia-Ruiz}}, \bibinfo {author} {\bibfnamefont {J.}~\bibnamefont {Billaud}},
  \bibinfo {author} {\bibfnamefont {J.~G.}\ \bibnamefont {Lozano}}, \bibinfo
  {author} {\bibfnamefont {R.~A.}\ \bibnamefont {House}}, \bibinfo {author}
  {\bibfnamefont {L.~C.}\ \bibnamefont {Gallington}}, \bibinfo {author}
  {\bibfnamefont {T.}~\bibnamefont {Ericsson}}, \bibinfo {author}
  {\bibfnamefont {L.}~\bibnamefont {H{\"a}ggstr{\"o}m}}, \bibinfo {author}
  {\bibfnamefont {M.~R.}\ \bibnamefont {Roberts}}, \bibinfo {author}
  {\bibfnamefont {U.}~\bibnamefont {Maitra}},\ and\ \bibinfo {author}
  {\bibfnamefont {P.~G.}\ \bibnamefont {Bruce}},\ }\href
  {https://doi.org/10.1039/C8EE02991A} {\bibfield  {journal} {\bibinfo
  {journal} {Energy {\&} Environmental Science}\ }\textbf {\bibinfo {volume}
  {12}},\ \bibinfo {pages} {2223} (\bibinfo {year} {2019})}\BibitemShut
  {NoStop}%
\bibitem [{\citenamefont {Usler}\ \emph {et~al.}(2023)\citenamefont {Usler},
  \citenamefont {Kemp}, \citenamefont {Bonkowski},\ and\ \citenamefont
  {de~Souza}}]{Usler.2023}%
  \BibitemOpen
  \bibfield  {author} {\bibinfo {author} {\bibfnamefont {A.~L.}\ \bibnamefont
  {Usler}}, \bibinfo {author} {\bibfnamefont {D.}~\bibnamefont {Kemp}},
  \bibinfo {author} {\bibfnamefont {A.}~\bibnamefont {Bonkowski}},\ and\
  \bibinfo {author} {\bibfnamefont {R.~A.}\ \bibnamefont {de~Souza}},\ }\href
  {https://doi.org/10.1002/jcc.27090} {\bibfield  {journal} {\bibinfo
  {journal} {Journal of computational chemistry}\ }\textbf {\bibinfo {volume}
  {44}},\ \bibinfo {pages} {1347} (\bibinfo {year} {2023})}\BibitemShut
  {NoStop}%
\bibitem [{\citenamefont {Mo}\ \emph {et~al.}(2014)\citenamefont {Mo},
  \citenamefont {Ong},\ and\ \citenamefont {Ceder}}]{Mo.2014}%
  \BibitemOpen
  \bibfield  {author} {\bibinfo {author} {\bibfnamefont {Y.}~\bibnamefont
  {Mo}}, \bibinfo {author} {\bibfnamefont {S.~P.}\ \bibnamefont {Ong}},\ and\
  \bibinfo {author} {\bibfnamefont {G.}~\bibnamefont {Ceder}},\ }\href
  {https://doi.org/10.1021/cm501563f} {\bibfield  {journal} {\bibinfo
  {journal} {Chemistry of Materials}\ }\textbf {\bibinfo {volume} {26}},\
  \bibinfo {pages} {5208} (\bibinfo {year} {2014})}\BibitemShut {NoStop}%
\bibitem [{\citenamefont {Katcho}\ \emph {et~al.}(2017)\citenamefont {Katcho},
  \citenamefont {Carrasco}, \citenamefont {Saurel}, \citenamefont {Gonzalo},
  \citenamefont {Han}, \citenamefont {Aguesse},\ and\ \citenamefont
  {Rojo}}]{Katcho.2017}%
  \BibitemOpen
  \bibfield  {author} {\bibinfo {author} {\bibfnamefont {N.~A.}\ \bibnamefont
  {Katcho}}, \bibinfo {author} {\bibfnamefont {J.}~\bibnamefont {Carrasco}},
  \bibinfo {author} {\bibfnamefont {D.}~\bibnamefont {Saurel}}, \bibinfo
  {author} {\bibfnamefont {E.}~\bibnamefont {Gonzalo}}, \bibinfo {author}
  {\bibfnamefont {M.}~\bibnamefont {Han}}, \bibinfo {author} {\bibfnamefont
  {F.}~\bibnamefont {Aguesse}},\ and\ \bibinfo {author} {\bibfnamefont
  {T.}~\bibnamefont {Rojo}},\ }\bibfield  {journal} {\bibinfo  {journal}
  {Advanced Energy Materials}\ }\textbf {\bibinfo {volume} {7}},\ \href
  {https://doi.org/10.1002/aenm.201601477} {10.1002/aenm.201601477} (\bibinfo
  {year} {2017})\BibitemShut {NoStop}%
\bibitem [{\citenamefont {Shu}\ and\ \citenamefont {Chou}(2008)}]{Shu.2008}%
  \BibitemOpen
  \bibfield  {author} {\bibinfo {author} {\bibfnamefont {G.~J.}\ \bibnamefont
  {Shu}}\ and\ \bibinfo {author} {\bibfnamefont {F.~C.}\ \bibnamefont {Chou}},\
  }\bibfield  {journal} {\bibinfo  {journal} {Physical Review B}\ }\textbf
  {\bibinfo {volume} {78}},\ \href {https://doi.org/10.1103/PhysRevB.78.052101}
  {10.1103/PhysRevB.78.052101} (\bibinfo {year} {2008})\BibitemShut {NoStop}%
\bibitem [{\citenamefont {Tatara}\ \emph {et~al.}(2025)\citenamefont {Tatara},
  \citenamefont {Igarashi}, \citenamefont {Nakayama}, \citenamefont {Hosaka},
  \citenamefont {Ohishi}, \citenamefont {Umegaki}, \citenamefont {Nakamura},
  \citenamefont {Koda}, \citenamefont {Ohta}, \citenamefont {Palm},
  \citenamefont {M{\aa}nsson}, \citenamefont {Kim}, \citenamefont {Kubota},
  \citenamefont {Sugiyama},\ and\ \citenamefont {Komaba}}]{Tatara.2025}%
  \BibitemOpen
  \bibfield  {author} {\bibinfo {author} {\bibfnamefont {R.}~\bibnamefont
  {Tatara}}, \bibinfo {author} {\bibfnamefont {D.}~\bibnamefont {Igarashi}},
  \bibinfo {author} {\bibfnamefont {M.}~\bibnamefont {Nakayama}}, \bibinfo
  {author} {\bibfnamefont {T.}~\bibnamefont {Hosaka}}, \bibinfo {author}
  {\bibfnamefont {K.}~\bibnamefont {Ohishi}}, \bibinfo {author} {\bibfnamefont
  {I.}~\bibnamefont {Umegaki}}, \bibinfo {author} {\bibfnamefont {J.~G.}\
  \bibnamefont {Nakamura}}, \bibinfo {author} {\bibfnamefont {A.}~\bibnamefont
  {Koda}}, \bibinfo {author} {\bibfnamefont {H.}~\bibnamefont {Ohta}}, \bibinfo
  {author} {\bibfnamefont {R.}~\bibnamefont {Palm}}, \bibinfo {author}
  {\bibfnamefont {M.}~\bibnamefont {M{\aa}nsson}}, \bibinfo {author}
  {\bibfnamefont {E.~J.}\ \bibnamefont {Kim}}, \bibinfo {author} {\bibfnamefont
  {K.}~\bibnamefont {Kubota}}, \bibinfo {author} {\bibfnamefont
  {J.}~\bibnamefont {Sugiyama}},\ and\ \bibinfo {author} {\bibfnamefont
  {S.}~\bibnamefont {Komaba}},\ }\bibfield  {journal} {\bibinfo  {journal}
  {Chemical science}\ }\href {https://doi.org/10.1039/d5sc03394b}
  {10.1039/d5sc03394b} (\bibinfo {year} {2025})\BibitemShut {NoStop}%
\bibitem [{\citenamefont {Ohishi}\ \emph {et~al.}(2023)\citenamefont {Ohishi},
  \citenamefont {Igarashi}, \citenamefont {Tatara}, \citenamefont {Umegaki},
  \citenamefont {Nakamura}, \citenamefont {Koda}, \citenamefont {M{\aa}nsson},
  \citenamefont {Komaba},\ and\ \citenamefont {Sugiyama}}]{Ohishi.2023}%
  \BibitemOpen
  \bibfield  {author} {\bibinfo {author} {\bibfnamefont {K.}~\bibnamefont
  {Ohishi}}, \bibinfo {author} {\bibfnamefont {D.}~\bibnamefont {Igarashi}},
  \bibinfo {author} {\bibfnamefont {R.}~\bibnamefont {Tatara}}, \bibinfo
  {author} {\bibfnamefont {I.}~\bibnamefont {Umegaki}}, \bibinfo {author}
  {\bibfnamefont {J.~G.}\ \bibnamefont {Nakamura}}, \bibinfo {author}
  {\bibfnamefont {A.}~\bibnamefont {Koda}}, \bibinfo {author} {\bibfnamefont
  {M.}~\bibnamefont {M{\aa}nsson}}, \bibinfo {author} {\bibfnamefont
  {S.}~\bibnamefont {Komaba}},\ and\ \bibinfo {author} {\bibfnamefont
  {J.}~\bibnamefont {Sugiyama}},\ }\href
  {https://doi.org/10.1021/acsaem.3c01197} {\bibfield  {journal} {\bibinfo
  {journal} {ACS Applied Energy Materials}\ }\textbf {\bibinfo {volume} {6}},\
  \bibinfo {pages} {8111} (\bibinfo {year} {2023})}\BibitemShut {NoStop}%
\bibitem [{\citenamefont {Shibata}\ \emph {et~al.}(2013)\citenamefont
  {Shibata}, \citenamefont {Kobayashi},\ and\ \citenamefont
  {Moritomo}}]{Shibata.2013}%
  \BibitemOpen
  \bibfield  {author} {\bibinfo {author} {\bibfnamefont {T.}~\bibnamefont
  {Shibata}}, \bibinfo {author} {\bibfnamefont {W.}~\bibnamefont {Kobayashi}},\
  and\ \bibinfo {author} {\bibfnamefont {Y.}~\bibnamefont {Moritomo}},\ }\href
  {https://doi.org/10.7567/APEX.6.097101} {\bibfield  {journal} {\bibinfo
  {journal} {Applied Physics Express}\ }\textbf {\bibinfo {volume} {6}},\
  \bibinfo {pages} {097101} (\bibinfo {year} {2013})}\BibitemShut {NoStop}%
\bibitem [{\citenamefont {Th{\"o}rnig}(2021)}]{Thornig.2021}%
  \BibitemOpen
  \bibfield  {author} {\bibinfo {author} {\bibfnamefont {P.}~\bibnamefont
  {Th{\"o}rnig}},\ }\href {https://doi.org/10.17815/jlsrf-7-182} {\bibfield
  {journal} {\bibinfo  {journal} {Journal of large-scale research facilities
  JLSRF}\ }\textbf {\bibinfo {volume} {7}},\ \bibinfo {pages} {A182} (\bibinfo
  {year} {2021})}\BibitemShut {NoStop}%
\end{thebibliography}

%

\end{document}